\documentclass[prd, twocolumn, superscriptaddress]{revtex4-2}
\usepackage{amsmath}
\usepackage{amssymb}

\usepackage{bm}

\usepackage{tikz}

\usepackage{array}

\allowdisplaybreaks[1]

\begin{document}
\title{Minkowski vacuum in Rindler spacetime and Unruh thermal state for Dirac fields}

\author{Riccardo Falcone}
\affiliation{Department of Physics, University of Sapienza, Piazzale Aldo Moro 5, 00185 Rome, Italy}

\author{Claudio Conti}
\affiliation{Department of Physics, University of Sapienza, Piazzale Aldo Moro 5, 00185 Rome, Italy}
\affiliation{Institute for Complex Systems (ISC-CNR), Department of Physics, University Sapienza, Piazzale Aldo Moro 2, 00185, Rome, Italy}
\affiliation{Research Center Enrico Fermi, Via Panisperna 89a, 00184 Rome, Italy}

\begin{abstract}
We consider a free Dirac field in flat spacetime and we derive the representation of the Minkowski vacuum as an element of the Rindler-Fock space. We also compute the statistical operator obtained by tracing away the left wedge. We detail the resulting thermal state for fermionic particles.
\end{abstract}

\maketitle

\section{Introduction}

The Unruh effect is the prediction that an accelerated observer detects Rindler particles in the Minkowski vacuum \cite{PhysRevD.7.2850, Davies:1974th, PhysRevD.14.870}. The phenomenon was originally studied in the context of scalar fields. Scalar particles in the accelerated frame are expected to follow the bosonic thermal distribution $(e^{\beta \Omega}-1)^{-1}$, where $\hbar \Omega$ is the particles energy and $\beta = 2 \pi / (c a)$ is inversely proportional to the acceleration of the observer $c^2 a$.

More recent works considered Dirac fields \cite{Oriti, PhysRevD.103.125005}. The result is a fermionic thermal distribution $(e^{\beta \Omega}+1)^{-1}$ for Rindler-Dirac particles in the Minkowski vacuum. Despite these investigations, a complete description of the Minkowski vacuum in terms of Dirac Rindler-Fock states is missing, as far as we know.

Here, we follow the algebraic approach \cite{wald1994quantum} to relate states of the inertial frame to the accelerated frame. We show how a Rindler-Fock representative for the Minkowski vacuum exists. Notice that the algebraic approach does not always guarantee the possibility to map the Fock space of one frame to the Fock space of another frame. The approach only guarantees the equivalence between mean values of states up to an arbitrarily large precision and for finite sets of operators. 

We compute the Bogoliubov transformation relating the Minkowski operators to the Rindler operators and we derive the Minkowski vacuum as an element of the Rindler-Fock space. We, hence, use the Bogoliubov coefficients to give a complete description of the Minkowski vacuum in the Rindler spacetime.

We obtain different Rindler-Fock representations depending of the choice for the spin basis in each wedge. The dependence of the spin basis is due to the presence of a spin coupling between modes of opposite wedges.

We also derive the statistical operator describing the Minkowski vacuum seen by the accelerated observer with positive acceleration. We compute the partial trace with respect to the left wedge by adopting a many-body approach for Dirac particles. The result is a fermionic thermal state that completely describes the Minkowski vacuum in the right Rindler spacetime.

The paper is organized as follows. In Sec.~\ref{Minkowski_Dirac_modes} we give a brief review of the Dirac field in Minkowski spacetime. In Sec.~\ref{Rindler_Dirac_modes} we consider the Rindler spacetime and derive the positive and negative frequency solutions of the Rindler-Dirac equation. In Sec.~\ref{Bogoliubov_transformation}, we compute the Bogoliubov transformations relating Minkowski to Rindler operators. The Bogoliubov coefficients are then used in Sec.~\ref{Minkowski_vacuum_in_Rindler_spacetime} to give the representation of the Minkowski vacuum in the Rindler spacetime. We compute the partial trace with respect to the left wedge and obtain the fermionic thermal state in Sec.~\ref{Unruh_effect_for_Dirac_fields}. In Sec.~\ref{Basis_choice} we discuss the dependence of the results with respect to the spin basis choice. Conclusions are drawn in Sec.~\ref{Conclusions}. Proofs related to the Bessel functions appearing in the Rindler-Dirac modes are provided in the Appendix \ref{Bessel_functions}.

\section{Minkowski-Dirac modes} \label{Minkowski_Dirac_modes}

In this section we give a brief review of free Dirac fields in Minkowski spacetime $(t,\vec{x})$. We derive the Minkowski-Dirac modes as orthonormal positive and negative frequency solutions of the Dirac equation. We consider modes with defined momentum, but we will not choose any particular basis for the spin degrees of freedom.

The Dirac field in Minkowski spacetime $\hat{\psi}(t,\vec{x})$ is described by the Dirac equation, that reads
\begin{equation} \label{Dirac}
\left( i c \gamma^\mu \partial_\mu  - \frac{m c^2}{\hbar} \right) \hat{\psi} = 0, 
\end{equation}
where $c$ is the speed of light, $m$ is the mass of the Dirac field,
\begin{align}\label{gamma_matrix}
& \gamma^0 = \frac{1}{c} \begin{pmatrix}
\mathbb{I} &0 \\
0 &-\mathbb{I}
\end{pmatrix}, & \gamma^i = \begin{pmatrix}
0 &\sigma^i \\
-\sigma^i &0
\end{pmatrix}
\end{align}
are gamma matrices and
\begin{align}\label{Pauli}
& \sigma^1 = \begin{pmatrix}
0 & 1 \\
1 & 0
\end{pmatrix}, & & \sigma^2 = \begin{pmatrix}
0 & -i \\
i & 0
\end{pmatrix}, & & \sigma^3 = \begin{pmatrix}
1 & 0 \\
0 & -1
\end{pmatrix}
\end{align}
are Pauli matrices. The following identities for gamma matrices will be used throughout the paper
\begin{subequations} \label{gamma_matrices_identities}
\begin{equation}\label{gamma_matrices_anticommutating_rule}
\{ \gamma^\mu, \gamma^\nu \} = -2 \eta^{\mu\nu},
\end{equation}
\begin{align}\label{gamma_matrices_hermitianity}
& (\gamma^0)^\dagger = \gamma^0, & (\gamma^i)^\dagger = -\gamma^i,
\end{align}
\end{subequations}
where
\begin{equation}
\eta^{\mu \nu} = \text{diag}\left(-\frac{1}{c^2}, 1, 1, 1\right)
\end{equation}
is the Minkowski metric.

The solution of the Dirac equation (\ref{Dirac}) is
\begin{equation} \label{free_Dirac_field}
\hat{\psi}(t,\vec{x}) = \sum_{s=1}^2 \int_{\mathbb{R}^3} d^3 k \left[ u_s(\vec{k},t,\vec{x}) \hat{c}_s(\vec{k}) + v_s(\vec{k},t,\vec{x}) \hat{d}_s^\dagger(\vec{k}) \right],
\end{equation}
with $\hat{c}_s(\vec{k})$ and $\hat{d}_s(\vec{k})$ as annihilation operators for the particle and antiparticle with momentum $\vec{k}$ and spin number $s$ and with $u_s(\vec{k})$ and $v_s(\vec{k})$ as orthonormal positive and negative frequency modes having the following form
\begin{subequations}\label{free_Dirac_field_modes}
\begin{align}
& u_s(\vec{k},t,\vec{x}) = (2\pi)^{-3/2} e^{ -i\omega(\vec{k})t + i\vec{k} \cdot \vec{x} }  \tilde{u}_s(\vec{k}), \label{u} \\
& v_s(\vec{k},t,\vec{x}) = (2\pi)^{-3/2} e^{ i\omega(\vec{k})t - i\vec{k} \cdot \vec{x} }  \tilde{v}_s(\vec{k}), \label{v}
\end{align}
\end{subequations}
with
\begin{equation}\label{omega_k}
\omega (\vec{k}) = \sqrt{ \left( \frac{m c^2}{\hbar} \right)^2 + c^2|\vec{k}|^2 }
\end{equation}
as the frequency of each mode.

The modes $u_s(\vec{k})$ and $v_s(\vec{k})$ are solutions of the Dirac equation (\ref{Dirac}). Their orthonormality condition reads
\begin{subequations}\label{D_scalar_product_orthonormality_u_v}
\begin{align}
& ( u_s(\vec{k}), u_{s'}(\vec{k}') )_\text{M} = \delta_{ss'} \delta^3(\vec{k}-\vec{k}'),\label{D_scalar_product_orthonormality_u_v_a}\\
 & ( v_s(\vec{k}), v_{s'}(\vec{k}') )_\text{M} = \delta_{ss'} \delta^3(\vec{k}-\vec{k}'),\\
  & ( u_s(\vec{k}), v_{s'}(\vec{k}') )_\text{M} = 0,
\end{align}
\end{subequations}
where
\begin{equation}\label{Dirac_scalar_product}
( \psi, \psi' )_\text{M} = \int_{\mathbb{R}^3} d^3x \psi^\dagger(t,\vec{x}) \psi'(t,\vec{x})
\end{equation}
is the Minkowski-Dirac product.

The functions $\tilde{u}_s(\vec{k})$ and $\tilde{v}_s(\vec{k})$ are solutions of the following equations 
\begin{subequations} \label{Dirac_uv_tilde}
\begin{align}
& \left[ \omega(\vec{k}) \gamma^0 - k_i \gamma^i - \frac{mc}{\hbar} \right] \tilde{u}_s(\vec{k}) = 0,\label{Dirac_u_tilde}\\
& \left[ \omega(\vec{k}) \gamma^0 - k_i \gamma^i + \frac{mc}{\hbar} \right] \tilde{v}_s(\vec{k}) = 0,\label{Dirac_v_tilde}\\
& \tilde{u}^\dagger_s(\vec{k}) \tilde{u}_{s'}(\vec{k}) = \delta_{ss'},\label{orthonormality_u_tilde}\\
& \tilde{v}^\dagger_s(\vec{k}) \tilde{v}_{s'}(\vec{k}) = \delta_{ss'},\label{orthonormality_v_tilde}\\
& \tilde{u}^\dagger_s(\vec{k}) \tilde{v}_{s'}(-\vec{k}) = 0.\label{orthonormality_uv_tilde}
\end{align}
\end{subequations}
One can use Eq.~(\ref{free_Dirac_field_modes}) and the fact that $u_s(\vec{k})$ and $v_s(\vec{k})$ are solutions of Eq.~(\ref{Dirac}) to obtain Eqs.~(\ref{Dirac_u_tilde}) and (\ref{Dirac_v_tilde}). The orthonormality conditions (\ref{orthonormality_u_tilde}), (\ref{orthonormality_v_tilde}) and (\ref{orthonormality_uv_tilde}), instead, can be checked by plugging Eq.~(\ref{free_Dirac_field_modes}) in Eq.~(\ref{D_scalar_product_orthonormality_u_v}) and using Eq.~(\ref{Dirac_scalar_product}).

The index $s$ is associated to the two independent spin degrees of freedom. One can consider any couple of solutions of Eq.~(\ref{Dirac_uv_tilde}) and associate each solution to either $s=1$ or $s=2$. This freedom is due to the arbitrary definition of the spin basis for positive and negative frequency modes.

A possible basis is given by particles with defined spin along one direction. For instance, states with spin up and down with respect to $z$ are such that in the particle comoving frame (i.e., by performing a Lorentz boost with opposite momentum $-\vec{k}$) the representative has only one spinorial component different from zero. A basis of modes with the same property is not available in the Rindler spacetime. Indeed, the translational symmetry with respect to the direction of the acceleration is absent, and particles with defined energy do not have defined momentum component along such a direction. Hence, no Lorentz boost leads to the comoving frame of these particles. For this reason, Rindler-Dirac modes with defined frequency and spin cannot be considered. 

Since our aim is to relate Minkowski modes with Rindler modes, there is no reason to prefer the basis with defined spin. Hereafter, we consider the general solutions of Eq.~(\ref{Dirac_uv_tilde}) and we do not choose any particular basis for the spin degrees of freedom.

\section{Rindler-Dirac modes}\label{Rindler_Dirac_modes}

In the previous sections we considered a Minkowski spacetime $(t,\vec{x})$ and studied the Minkowski-Dirac modes $u_s(\vec{k})$ and $v_s(\vec{k})$ as positive and negative frequency solutions of the Dirac equation. Here, instead, we consider Rindler coordinates $(T,\vec{X})$ for the left and the right wedge, defined as follows
\begin{align}\label{Rindler_coordinates_transformation}
& t = t_\nu(T,\vec{X}), & \vec{x} = \vec{x}_\nu(T,\vec{X}),
\end{align}
where $\nu \in \{ \text{L}, \text{R} \}$ is the variable associated to the left (L) and the right (R) wedges and $t_\nu(T,\vec{X})$ and $\vec{x}_\nu(T,\vec{X})$ are the coordinate transformation from the Rindler to the Minkowski frame. By assuming that the acceleration of the Rindler observer is along the direction of $z$, one may write
\begin{align}\label{Rindler_coordinates_transformation_2}
& z = z_\nu(T,\vec{X}), & \vec{x}_\perp = \vec{X}_\perp,
\end{align}
where $\vec{x}_\perp = (x,y)$ and $\vec{X}_\perp = (X,Y)$ are the transverse coordinates in each frame. The functions $t_\nu(T,\vec{X})$ and $z_\nu(T,\vec{X})$ appearing in Eqs.~(\ref{Rindler_coordinates_transformation}) and (\ref{Rindler_coordinates_transformation_2}) read
\begin{subequations}
\begin{align}
& t_\nu(T,\vec{X}) = \frac{e^{s_\nu a Z}}{c a} \sinh(c a T), \\
 & z_\nu(T,\vec{X}) = s_\nu \frac{e^{s_\nu a Z}}{a} \cosh(c a T),
\end{align}
\end{subequations}
where $c^2 a>0$ is the acceleration of the right Rindler observer and $s_\nu$ is such that $s_\text{L} = -1$ and $s_\text{R} = 1$. The left and the right wedges are defined by $z<c|t|$ and $z>c|t|$, respectively.

We study the Dirac field in Rindler coordinates $\hat{\Psi}_\nu (T,\vec{X})$ defined by the Rindler-Dirac equation, which reads (see, for instance, Refs.~\cite{Oriti, PhysRevD.103.125005, falcone2022non})
\begin{align} \label{Dirac_Rindler}
& \left[  e^{-s_\nu aZ} \left( i c \gamma^0 \partial_0 + s_\nu i  \frac{c a}{2} \gamma^3 +  i c \gamma^3 \partial_3 \right) \right. \nonumber \\
& \left. + i c \gamma^1 \partial_1 + i c \gamma^2 \partial_2 - \frac{m c^2}{\hbar}  \right] \hat{\Psi}_\nu = 0.
\end{align}
We derive the orthonormal positive and negative frequency modes $U_{\nu s}(\Omega,\vec{K}_\perp)$ and $V_{\nu s}(\Omega,\vec{K}_\perp)$ that are solutions of Eq.~(\ref{Dirac_Rindler}), have the following form
\begin{subequations}\label{UV_UV_tilde}
\begin{align}
& U_{\nu s}(\Omega,\vec{K}_\perp,T,\vec{X}) = e^{i \vec{K}_\perp \cdot \vec{X}_\perp - i \Omega T} \tilde{U}_{\nu s}(\Omega,\vec{K}_\perp,Z),\label{U_U_tilde} \\
& V_{\nu s}(\Omega,\vec{K}_\perp,T,\vec{X}) = e^{-i \vec{K}_\perp \cdot \vec{X}_\perp + i \Omega T} \tilde{V}_{\nu s}(\Omega,\vec{K}_\perp,Z)\label{V_V_tilde}
\end{align}
\end{subequations}
and generate Dirac fields in Rindler spacetime as follows
\begin{align} \label{free_Dirac_field_Rindler}
\hat{\Psi}_\nu(T,\vec{X}) = & \sum_{s=1}^2 \int_0^\infty d\Omega \int_{\mathbb{R}^2} d^2 K_\perp  \nonumber \\
& \times \left[ U_{\nu s}(\Omega,\vec{K}_\perp,T,\vec{X}) \hat{C}_{\nu s}(\Omega,\vec{K}_\perp) \right. \nonumber \\
& \left. + V_{\nu s}(\Omega,\vec{K}_\perp,T,\vec{X}) \hat{D}_{\nu s}^\dagger(\Omega,\vec{K}_\perp) \right].
\end{align}
The orthonormality condition for such modes reads
\begin{subequations}\label{D_scalar_product_orthonormality_U_V}
\begin{align}
( U_{\nu s}(\Omega,\vec{K}_\perp), U_{\nu s'}(\Omega',\vec{K}'_\perp) )_\nu = & \delta_{ss'} \delta(\Omega-\Omega') \nonumber \\ & \times \delta^2(\vec{K}_\perp-\vec{K}'_\perp),\label{D_scalar_product_orthonormality_U}\\
 ( V_{\nu s}(\Omega,\vec{K}_\perp), V_{\nu s'}(\Omega',\vec{K}'_\perp) )_\nu = & \delta_{ss'} \delta(\Omega-\Omega') \nonumber \\ & \times \delta^2(\vec{K}_\perp-\vec{K}'_\perp),\label{D_scalar_product_orthonormality_V}\\
  (U_{\nu s}(\Omega,\vec{K}_\perp), V_{\nu s'}(\Omega',\vec{K}'_\perp) )_\nu = & 0,
\end{align}
\end{subequations}
with
\begin{equation}\label{Dirac_Rindler_scalar_product}
( \Psi, \Psi' )_\nu = \int_{\mathbb{R}^3} d^3X e^{s_\nu a Z} \Psi^\dagger(T,\vec{X}) \Psi'(T,\vec{X})
\end{equation}
as Rindler-Dirac product.

The operators $\hat{C}^\dagger_{\nu s}(\Omega,\vec{K}_\perp)$ and $\hat{D}^\dagger_{\nu s}(\Omega,\vec{K}_\perp)$ appearing in Eq.~(\ref{free_Dirac_field_Rindler}) create particles and antiparticles of the $\nu$ wedge with spin number $s$, frequency $\Omega$ and transverse momentum $\vec{K}_\perp$ and satisfy the following anticommutation rules
\begin{subequations}\label{anticommutating_rules}
\begin{align}
\{ \hat{C}_{\nu s}(\Omega,\vec{K}_\perp), \hat{C}^\dagger_{\nu' s'}(\Omega',\vec{K}'_\perp) \}  = & \delta_{\nu \nu'} \delta_{ss'} \delta(\Omega-\Omega')\nonumber\\
& \times \delta^2(\vec{K}_\perp-\vec{K}'_\perp), \\
 \{ \hat{D}_{\nu s}(\Omega,\vec{K}_\perp), \hat{D}^\dagger_{\nu' s'}(\Omega',\vec{K}'_\perp) \}  = & \delta_{\nu \nu'} \delta_{ss'} \delta(\Omega-\Omega')\nonumber\\
& \times \delta^2(\vec{K}_\perp-\vec{K}'_\perp),  \\
 \{ \hat{C}_{\nu s}(\Omega,\vec{K}_\perp), \hat{C}_{\nu' s'}(\Omega',\vec{K}'_\perp) \}  = & 0, \label{anticommutating_rules_c} \\
 \{ \hat{D}_{\nu s}(\Omega,\vec{K}_\perp), \hat{D}_{\nu' s'}(\Omega',\vec{K}'_\perp) \}  = & 0, \label{anticommutating_rules_d}\\
  \{ \hat{C}_{\nu s}(\Omega,\vec{K}_\perp), \hat{D}_{\nu' s'}(\Omega',\vec{K}'_\perp) \}  = & 0, \label{anticommutating_rules_e}\\
  \{ \hat{C}_{\nu s}(\Omega,\vec{K}_\perp), \hat{D}^\dagger_{\nu' s'}(\Omega',\vec{K}'_\perp) \}  = & 0 \label{anticommutating_rules_f}.
\end{align}
\end{subequations}

By extending the definition of the variable $\Omega$ also for negative values, one may define the following function
\begin{equation} \label{W_UV}
W_{\nu s}(\Omega,\vec{K}_\perp,T,\vec{X}) = \begin{cases}
U_{\nu s}(\Omega,\vec{K}_\perp,T,\vec{X}) & \text{if } \Omega>0 \\
V_{\nu s}(-\Omega,-\vec{K}_\perp,T,\vec{X}) & \text{if } \Omega<0
\end{cases},
\end{equation}
that includes both positive and negative frequency solutions of the Rindler-Dirac equation (\ref{Dirac_Rindler}). Equation (\ref{free_Dirac_field_Rindler}) can now be written in the following equivalent ways
\begin{subequations}\label{free_Dirac_field_Rindler_2}
\begin{align} 
& \hat{\Psi}_\nu(T,\vec{X}) = \sum_{s=1}^2 \int_\mathbb{R} d\Omega \int_{\mathbb{R}^2} d^2 K_\perp W_{\nu s}(\Omega,\vec{K}_\perp,T,\vec{X})   \nonumber \\
& \times \left[ \theta(\Omega) \hat{C}_{\nu s}(\Omega,\vec{K}_\perp)+  \theta(-\Omega)\hat{D}_{\nu s}^\dagger(-\Omega,-\vec{K}_\perp) \right], \\
& \hat{\Psi}_\nu(T,\vec{X}) = \sum_{s=1}^2 \int_\mathbb{R} d\Omega \int_{\mathbb{R}^2} d^2 K_\perp W_{\nu s}(-\Omega,-\vec{K}_\perp,T,\vec{X})   \nonumber \\
& \times \left[ \theta(-\Omega) \hat{C}_{\nu s}(-\Omega,-\vec{K}_\perp)+  \theta(\Omega)\hat{D}_{\nu s}^\dagger(\Omega,\vec{K}_\perp) \right],
\end{align}
\end{subequations}
with $\theta$ as the Heaviside step function. The orthonormality condition (\ref{D_scalar_product_orthonormality_U_V}) with respect to the modes $W_{\nu s}(\Omega,\vec{K}_\perp)$ reads
\begin{align}\label{D_scalar_product_orthonormality_W}
( W_{\nu s}(\Omega,\vec{K}_\perp), W_{\nu s'}(\Omega',\vec{K}'_\perp) )_\nu = & \delta_{ss'} \delta(\Omega-\Omega') \nonumber \\ & \times \delta^2(\vec{K}_\perp-\vec{K}'_\perp).
\end{align}

Notice that Eq.~(\ref{W_UV}) is compatible with Eq.~(\ref{UV_UV_tilde}). Indeed, one may define the function $\tilde{W}_{\nu s}(\Omega,\vec{K}_\perp,Z)$ such that
\begin{equation} \label{UVW_UVW_tilde}
W_{\nu s}(\Omega,\vec{K}_\perp,T,\vec{X}) = e^{i \vec{K}_\perp \cdot \vec{X}_\perp - i \Omega T} \tilde{W}_{\nu s}(\Omega,\vec{K}_\perp,Z)
\end{equation}
and that
\begin{equation} \label{W_tilde_UV_tilde}
\tilde{W}_{\nu s}(\Omega,\vec{K}_\perp,Z) = \begin{cases}
\tilde{U}_{\nu s}(\Omega,\vec{K}_\perp,Z) & \text{if } \Omega>0 \\
\tilde{V}_{\nu s}(-\Omega,-\vec{K}_\perp,Z) & \text{if } \Omega<0
\end{cases}.
\end{equation}

The modes $W_{\nu s}(\Omega,\vec{K}_\perp)$ are solutions of the Rindler-Dirac equation (\ref{Dirac_Rindler}); hence, $\tilde{W}_{\nu s}(\Omega,\vec{K}_\perp,Z)$ satisfies the following differential equation
\begin{align} \label{Dirac_Rindler_2}
& \left[ e^{-s_\nu  aZ} \left( \Omega \gamma^0 + s_\nu i \frac{a}{2} \gamma^3 +  i \gamma^3 \partial_3  \right) \right. \nonumber \\
& \left. -  \left( K_1 \gamma^1 + K_2 \gamma^2 + \frac{m c}{\hbar} \right) \right] \tilde{W}_{\nu s}(\Omega,\vec{K}_\perp,Z) = 0.
\end{align}
By multiplying Eq.~(\ref{Dirac_Rindler_2}) with $c \gamma^0$ on the left and using Eq.~(\ref{gamma_matrices_anticommutating_rule}), one obtains
\begin{align} \label{Dirac_Rindler_3}
& \left[  e^{-s_\nu aZ} \left( \frac{\Omega}{c} + s_\nu i \frac{c a}{2} \gamma^0 \gamma^3 +  i c \gamma^0 \gamma^3 \partial_3 \right)  \right. \nonumber \\
& \left. - s_\nu i \kappa (\vec{K}_\perp) \mathfrak{G}_\nu (\vec{K}_\perp) \right] \tilde{W}_{\nu s}(\Omega,\vec{K}_\perp,Z) = 0,
\end{align}
with
\begin{equation}\label{Gamma}
\mathfrak{G}_\nu (\vec{K}_\perp) = - \frac{s_\nu i c} {\kappa (\vec{K}_\perp)} \gamma^0 \left( K_1 \gamma^1 + K_2 \gamma^2 + \frac{m c}{\hbar} \right),
\end{equation}
and
\begin{equation}\label{kappa_k_perp}
\kappa (\vec{K}_\perp) = \sqrt{ \left( \frac{m c}{\hbar} \right)^2 + |\vec{K}_\perp|^2 }.
\end{equation}

The spinor $\tilde{W}_{\nu s}(\Omega,\vec{K}_\perp,Z)$ can be decomposed into eigenvectors of $c \gamma^0 \gamma^3$ with eigenvalues $\pm 1$ by using the following projectors
\begin{equation}\label{P_pm}
P_\pm = \frac{1}{2} (1 \pm c \gamma^0 \gamma^3).
\end{equation}
The projected modes $\tilde{W}^\pm_{\nu s}(\Omega,\vec{K}_\perp,Z)$ are such that
\begin{subequations}
\begin{align}
& \tilde{W}_{\nu s}(\Omega,\vec{K}_\perp,Z) = \tilde{W}^+_{\nu s}(\Omega,\vec{K}_\perp,Z) + \tilde{W}^-_{\nu s}(\Omega,\vec{K}_\perp,Z), \label{U_U_p_U_m}\\
 & \tilde{W}^\pm_{\nu s}(\Omega,\vec{K}_\perp,Z) = P_\pm \tilde{W}_{\nu s}(\Omega,\vec{K}_\perp,Z), \label{U_pm}\\ 
& c\gamma^0 \gamma^3 \tilde{W}^\pm_{\nu s}(\Omega,\vec{K}_\perp,Z) = \pm \tilde{W}^\pm_{\nu s}(\Omega,\vec{K}_\perp,Z) \label{U_pm_eigenvectors}.
\end{align}
\end{subequations}
By using Eqs.~(\ref{gamma_matrices_anticommutating_rule}) and (\ref{Gamma}) one can prove that
\begin{equation}\label{gamma_03_gamma_012}
\gamma^0 \gamma^3 \mathfrak{G}_\nu (\vec{K}_\perp) = - \mathfrak{G}_\nu (\vec{K}_\perp) \gamma^0 \gamma^3.
\end{equation}
Hence, the projectors $P_\pm $ and the matrix $\mathfrak{G}_\nu (\vec{K}_\perp)$ are related by the following identity
\begin{equation}\label{P_mp_gamma_012}
P_\pm \mathfrak{G}_\nu (\vec{K}_\perp) = \mathfrak{G}_\nu (\vec{K}_\perp)  P_\mp,
\end{equation}
which can be proved by using Eqs.~(\ref{P_pm}) and (\ref{gamma_03_gamma_012}). By projecting Eq.~(\ref{Dirac_Rindler_3}) with respect to $P_\pm$ and using Eqs.~(\ref{U_pm}), (\ref{U_pm_eigenvectors}) and (\ref{P_mp_gamma_012}), one obtains the following coupled equations for $\tilde{W}^+_{\nu s}(\Omega,\vec{K}_\perp,Z)$ and $\tilde{W}^-_{\nu s}(\Omega,\vec{K}_\perp,Z)$
\begin{align} \label{Dirac_Rindler_4}
& e^{-s_\nu aZ} \left[ \frac{\Omega}{c} \pm i \left( s_\nu \frac{a}{2}  +  \partial_3 \right)  \right] \tilde{W}^\pm_{\nu s}(\Omega,\vec{K}_\perp,Z) \nonumber \\
= & s_\nu i \kappa (\vec{K}_\perp) \mathfrak{G}_\nu (\vec{K}_\perp) \tilde{W}^\mp_{\nu s}(\Omega,\vec{K}_\perp,Z).
\end{align}

Equation (\ref{Dirac_Rindler_4}) can be decoupled by applying $e^{-s_\nu aZ}[\Omega/c \mp i (s_\nu a/2 + \partial_3)]$ on the left, leading to
\begin{align} \label{Dirac_Rindler_5}
& e^{-s_\nu aZ} \left[ \frac{\Omega}{c} \mp i \left( s_\nu \frac{a}{2}  +  \partial_3 \right)  \right] \left\lbrace e^{-s_\nu aZ} \left[ \frac{\Omega}{c} \right. \right. \nonumber \\
& \left. \left. \pm i \left( s_\nu \frac{a}{2}  +  \partial_3 \right)  \right] \right\rbrace \tilde{W}^\pm_{\nu s}(\Omega,\vec{K}_\perp,Z) \nonumber \\
= & - \kappa^2 (\vec{K}_\perp)  \mathfrak{G}^2_\nu (\vec{K}_\perp)   \tilde{W}^\pm_{\nu s}(\Omega,\vec{K}_\perp,Z).
\end{align}
The derivative operator on left side of Eq.~(\ref{Dirac_Rindler_5}) can be computed in the following way
\begin{align} \label{Dirac_Rindler_5_left}
& \left[ \frac{\Omega}{c} \mp i \left( s_\nu \frac{a}{2}  +  \partial_3 \right)  \right] \left\lbrace e^{-s_\nu aZ} \left[ \frac{\Omega}{c} \pm i \left( s_\nu \frac{a}{2}  +  \partial_3 \right)  \right] \right\rbrace \nonumber \\
= & e^{-s_\nu aZ}  \left\lbrace \pm s_\nu i a \left[ \frac{\Omega}{c} \pm i \left( s_\nu \frac{a}{2}  +  \partial_3 \right)  \right] + \left( \frac{\Omega}{c} \right)^2 \right. \nonumber \\
& \left.+ \left( s_\nu \frac{a}{2}  +  \partial_3 \right)^2  \right\rbrace \nonumber \\
= & e^{-s_\nu aZ}  \left[ \pm s_\nu i a  \frac{\Omega}{c} - \frac{a^2}{4}   + \left( \frac{\Omega}{c} \right)^2  +  \partial^2_3    \right] \nonumber \\
= & e^{-s_\nu aZ}  \left[ \left( \frac{\Omega}{c} \pm s_\nu i \frac{a}{2}  \right)^2  +  \partial^2_3    \right].
\end{align}
The right side of Eq.~(\ref{Dirac_Rindler_5}), instead, can be computed by using Eqs.~(\ref{gamma_matrices_anticommutating_rule}), (\ref{Gamma}) and (\ref{kappa_k_perp}),
\begin{align}\label{Dirac_Rindler_5_right}
\mathfrak{G}^2_\nu (\vec{K}_\perp)= & - \frac{c^2} {\kappa^2 (\vec{K}_\perp)} \gamma^0 \left( K_1 \gamma^1 + K_2 \gamma^2 + \frac{m c}{\hbar} \right)  \nonumber \\
& \times \gamma^0 \left( K_1 \gamma^1 + K_2 \gamma^2 + \frac{m c}{\hbar} \right)  \nonumber \\
= &  - \frac{c^2} {\kappa^2 (\vec{K}_\perp)} \gamma^0 \gamma^0 \left( -K_1 \gamma^1 - K_2 \gamma^2 + \frac{m c}{\hbar} \right) \nonumber \\
& \times \left( K_1 \gamma^1 + K_2 \gamma^2 + \frac{m c}{\hbar} \right)  \nonumber \\
= & - \frac{1} {\kappa^2 (\vec{K}_\perp)} \left( -K_1 \gamma^1 - K_2 \gamma^2 + \frac{m c}{\hbar} \right) \nonumber \\
& \times \left( K_1 \gamma^1 + K_2 \gamma^2 + \frac{m c}{\hbar} \right)  \nonumber \\
= & - \frac{1} {\kappa^2 (\vec{K}_\perp)} \left[  -K_1^2 \gamma^1 \gamma^1 - K_2^2 \gamma^2 \gamma^2 + \left( \frac{m c}{\hbar} \right)^2 \right. \nonumber \\
&  \left. - K_1 K_2 \{\gamma^1, \gamma^2 \} \right] \nonumber \\
= & - \frac{1} {\kappa^2 (\vec{K}_\perp)} \left[  K_1^2 + K_2^2  + \left( \frac{m c}{\hbar} \right)^2 \right] \nonumber \\
= &  -1.
\end{align}
By using Eqs.~(\ref{Dirac_Rindler_5_left}) and (\ref{Dirac_Rindler_5_right}) in Eq.~(\ref{Dirac_Rindler_5}), one obtains the following differential equation
\begin{align} \label{Dirac_Rindler_6}
&  e^{-s_\nu 2aZ} \left[ \left( \frac{\Omega}{c} \pm s_\nu i \frac{a}{2} \right)^2 + \partial_3^2  \right] \tilde{W}^\pm_{\nu s}(\Omega,\vec{K}_\perp,Z) \nonumber \\
= & \kappa^2 (\vec{K}_\perp)  \tilde{W}^\pm_{\nu s}(\Omega,\vec{K}_\perp,Z).
\end{align}

The solutions of Eq.~(\ref{Dirac_Rindler_6}) that converge to $0$ for $s_\nu Z \rightarrow +\infty$ have the following form
\begin{equation}\label{U_pm_s_W_pm_s}
\tilde{W}^\pm_{\nu s}(\Omega,\vec{K}_\perp,Z) = \mathfrak{K} (\pm s_\nu \Omega,\vec{K}_\perp,s_\nu Z) \mathfrak{W}^\pm_{\nu s}(\Omega,\vec{K}_\perp),
\end{equation}
where
\begin{equation}\label{K_pm}
\mathfrak{K}(\Omega,\vec{K}_\perp,Z) = K_{i \Omega / (c a) - 1/2} \left( \kappa (\vec{K}_\perp) \frac{e^{aZ}}{a} \right)
\end{equation}
and $K_\zeta (\xi)$ is the modified Bessel function of the second kind. An integral representation for $K_\zeta (\xi)$ can be found in Appendix \ref{Bessel_functions}. Notice that Eq.~(\ref{Dirac_Rindler_6}) is a necessary but not sufficient condition for the modes $\tilde{W}^\pm_{\nu s}(\Omega,\vec{K}_\perp,Z)$. Indeed, Eq.~(\ref{Dirac_Rindler_6}) is a decoupled second order differential equation originated from the first order differential equation (\ref{Dirac_Rindler_4}). Hence, we now look for the spinor functions $\mathfrak{W}^\pm_{\nu s}(\Omega,\vec{K}_\perp)$ of Eq.~(\ref{U_pm_s_W_pm_s}) such that $\tilde{W}^\pm_{\nu s}(\Omega,\vec{K}_\perp,Z)$ satisfies Eq.~(\ref{Dirac_Rindler_4}).

The first order derivatives of $\tilde{W}^\pm_{\nu s}(\Omega,\vec{K}_\perp,Z)$ that appear in Eq.~(\ref{Dirac_Rindler_4}) can be computed by using the following recurrence relation for Bessel functions \cite{abramowitz1965handbook}
\begin{equation}\label{Bessel_derivative_recursive}
\partial_\xi K_\zeta (\xi) - \frac{\zeta}{\xi} K_\zeta (\xi) = -K_{\zeta + 1} (\xi)
\end{equation}
and the fact that $K_\zeta (\xi)$ is even with respect to the order $\zeta$, which means that
\begin{equation}\label{Bessel_derivative_recursive_2}
\partial_\xi K_\zeta (\xi) - \frac{\zeta}{\xi} K_\zeta (\xi) = -K_{-\zeta - 1} (\xi).
\end{equation}
By considering $\xi = \kappa (\vec{K}_\perp) \exp(s_\nu aZ)/a$ and $\zeta = \pm s_\nu i \Omega / (c a) - 1/2 $, one may write Eq.~(\ref{Bessel_derivative_recursive_2}) in terms of the functions $\mathfrak{K}(\Omega,\vec{K}_\perp,Z)$ [Eq.~(\ref{K_pm})] as follows
\begin{align}\label{K_recurrence}
&  \frac{e^{-s_\nu aZ}}{\kappa (\vec{K}_\perp)} \left(s_\nu \partial_3 \mp s_\nu  i \frac{\Omega}{c} + \frac{a}{2} \right) \mathfrak{K}(\pm s_\nu \Omega,\vec{K}_\perp,s_\nu Z)  \nonumber \\
= & - \mathfrak{K} (\mp s_\nu \Omega,\vec{K}_\perp,s_\nu Z) ,
\end{align}
which, multiplied with $ \pm s_\nu i \kappa (\vec{K}_\perp)$, reads
\begin{align}\label{K_recurrence_2}
&  e^{-s_\nu aZ} \left[\frac{\Omega}{c} \pm i \left( s_\nu \frac{a}{2} + \partial_3 \right) \right] \mathfrak{K}(\pm s_\nu \Omega,\vec{K}_\perp,s_\nu Z)  \nonumber \\
= & \mp s_\nu i \kappa (\vec{K}_\perp) \mathfrak{K} (\mp s_\nu \Omega,\vec{K}_\perp,s_\nu Z) .
\end{align}
By using Eqs.~(\ref{U_pm_s_W_pm_s}) and (\ref{K_recurrence_2}) in Eq.~(\ref{Dirac_Rindler_4}) one obtains the following linear equation for $\mathfrak{W}^\pm_{\nu s}(\Omega,\vec{K}_\perp)$
\begin{equation} \label{Dirac_Rindler_4_W}
 \mathfrak{W}^\pm_{\nu s}(\Omega,\vec{K}_\perp)= \mp  \mathfrak{G}_\nu (\vec{K}_\perp)  \mathfrak{W}^\mp_{\nu s}(\Omega,\vec{K}_\perp).
\end{equation}

The two equations appearing in Eq.~(\ref{Dirac_Rindler_4_W}) are equivalent. This can be proven by acting on Eq.~(\ref{Dirac_Rindler_4_W}) with $ \pm \mathfrak{G}_\nu (\vec{K}_\perp)$ and by using Eq.~(\ref{Dirac_Rindler_5_right}). Hence, one can consider a single spinor function $\mathfrak{W}_{\nu s}(\Omega,\vec{K}_\perp)$ such that
\begin{equation}\label{W_pm_W}
\mathfrak{W}^\pm_{\nu s}(\Omega,\vec{K}_\perp) = \left[ \mathfrak{G}_\nu(\vec{K}_\perp) \right]^{(1 \mp 1)/2}  \mathfrak{W}_{\nu s}(\Omega,\vec{K}_\perp).
\end{equation}

Notice that each spinor $\mathfrak{W}^\pm_{\nu s} (\Omega,\vec{K}_\perp)$ is eigenvector of $c \gamma^0 \gamma^3$ with eigenvalue $\pm 1$ [Eqs.~(\ref{U_pm_eigenvectors}) and (\ref{U_pm_s_W_pm_s})]. Hence the following identity must be considered together with Eq.~(\ref{W_pm_W})
\begin{equation}\label{W_pm_eigenvalues}
c\gamma^0 \gamma^3 \mathfrak{W}^\pm_{\nu s}(\Omega,\vec{K}_\perp) = \pm \mathfrak{W}^\pm_{\nu s}(\Omega,\vec{K}_\perp).
\end{equation}
Equations (\ref{W_pm_W}) and (\ref{W_pm_eigenvalues}) are outnumbered. Indeed, one may consider one of the two equations appearing in Eq.~(\ref{W_pm_eigenvalues}) and obtain the other by using Eq.~(\ref{W_pm_W}). For instance, by choosing $c\gamma^0 \gamma^3 \mathfrak{W}^+_{\nu s}(\Omega,\vec{K}_\perp) = \mathfrak{W}^+_{\nu s}(\Omega,\vec{K}_\perp)$, one can use Eq.~(\ref{gamma_03_gamma_012}) and (\ref{W_pm_W}) to prove that
\begin{align}
 c\gamma^0 \gamma^3 \mathfrak{W}^-_{\nu s}(\Omega,\vec{K}_\perp) = & c\gamma^0 \gamma^3 \mathfrak{G}_\nu(\vec{K}_\perp)  \mathfrak{W}_{\nu s}(\Omega,\vec{K}_\perp)\nonumber \\
= & c\gamma^0 \gamma^3 \mathfrak{G}_\nu(\vec{K}_\perp)  \mathfrak{W}^+_{\nu s}(\Omega,\vec{K}_\perp)\nonumber \\
= & - c \mathfrak{G}_\nu(\vec{K}_\perp) \gamma^0 \gamma^3  \mathfrak{W}^+_{\nu s}(\Omega,\vec{K}_\perp)\nonumber \\
= & - \mathfrak{G}_\nu(\vec{K}_\perp)  \mathfrak{W}^+_{\nu s}(\Omega,\vec{K}_\perp) \nonumber \\
= & - \mathfrak{G}_\nu(\vec{K}_\perp)  \mathfrak{W}_{\nu s}(\Omega,\vec{K}_\perp) \nonumber \\
= & - \mathfrak{W}^-_{\nu s}(\Omega,\vec{K}_\perp).
\end{align}
Both equations appearing in Eq.~(\ref{W_pm_eigenvalues}) are equivalent to the following single equation
\begin{equation}\label{W_p_eigenvalues}
c\gamma^0 \gamma^3 \mathfrak{W}_{\nu s}(\Omega,\vec{K}_\perp) = \mathfrak{W}_{\nu s}(\Omega,\vec{K}_\perp).
\end{equation}
 
The third identity defining $\mathfrak{W}_{\nu s} (\Omega,\vec{K}_\perp)$ comes from the orthonormality condition (\ref{D_scalar_product_orthonormality_W}). The product $( W_{\nu s}(\Omega,\vec{K}_\perp), W_{\nu {s'}}(\Omega',\vec{K}'_\perp) )_\nu$ can be computed by using Eqs.~(\ref{Dirac_Rindler_scalar_product}), (\ref{UVW_UVW_tilde}), (\ref{U_U_p_U_m}), (\ref{U_pm_s_W_pm_s}) and the orthogonality condition between eigenstates of $c \gamma^0 \gamma^3$ with different eigenvalues. Explicitly, the product reads
\begin{align} \label{W_R_W_R_product}
& ( W_{\nu s}(\Omega,\vec{K}_\perp), W_{\nu {s'}}(\Omega',\vec{K}'_\perp) )_\nu  = e^{ i( \Omega - \Omega') T}  \nonumber \\
 & \times \sum_{\sigma = \pm} \int_{\mathbb{R}^3} d^3X e^{s_\nu aZ} e^{i (\vec{K}'_\perp-\vec{K}_\perp) \cdot \vec{X}_\perp}  \mathfrak{K}^*(\sigma s_\nu \Omega,\vec{K}_\perp,s_\nu Z)  \nonumber \\
 & \times   \mathfrak{K}(\sigma s_\nu \Omega',\vec{K}_\perp',s_\nu Z) \left[ \mathfrak{W}^\sigma_{\nu s} (\Omega,\vec{K}_\perp) \right]^\dagger \mathfrak{W}^\sigma_{\nu s'} (\Omega',\vec{K}_\perp').
\end{align}

By using Eqs.~(\ref{gamma_matrices_identities}) and (\ref{Gamma}), one can prove that $\mathfrak{G}_\nu(\vec{K}_\perp)$ is antihermitian
\begin{align}\label{G_antihermitian}
 \mathfrak{G}^\dagger_\nu(\vec{K}_\perp) = & \frac{s_\nu i c} {\kappa (\vec{K}_\perp)} \left[ \gamma^0 \left( K_1 \gamma^1 + K_2 \gamma^2 + \frac{m c}{\hbar} \right) \right]^\dagger \nonumber \\
= & \frac{s_\nu i c} {\kappa (\vec{K}_\perp)} \left(-K_1 \gamma^1 - K_2 \gamma^2 + \frac{m c}{\hbar} \right) \gamma^0 \nonumber \\
= & \frac{s_\nu i c} {\kappa (\vec{K}_\perp)} \gamma^0 \left(K_1 \gamma^1 + K_2 \gamma^2 + \frac{m c}{\hbar} \right) \nonumber \\
= & - \mathfrak{G}_\nu(\vec{K}_\perp).
\end{align}
Equations (\ref{Dirac_Rindler_5_right}) and (\ref{G_antihermitian}) imply that $\mathfrak{G}_\nu(\vec{K}_\perp)$ is also unitary
\begin{equation}\label{G_unitarity}
\mathfrak{G}^\dagger_\nu(\vec{K}_\perp) \mathfrak{G}_\nu(\vec{K}_\perp) = 1.
\end{equation}
By using Eqs.~(\ref{W_pm_W}) and (\ref{G_unitarity}) one can prove that for any $\sigma=\pm$,
\begin{align}
 & \left[ \mathfrak{W}^\sigma_{\nu s} (\Omega,\vec{K}_\perp) \right]^\dagger \mathfrak{W}^\sigma_{\nu s'} (\Omega',\vec{K}_\perp') \nonumber \\
  = & \mathfrak{W}^\dagger_{\nu s} (\Omega,\vec{K}_\perp) \mathfrak{W}_{\nu s'} (\Omega',\vec{K}_\perp'),
\end{align}
which means that Eq.~(\ref{W_R_W_R_product}) reads
\begin{align} \label{W_R_W_R_product_2}
& ( W_{\nu s}(\Omega,\vec{K}_\perp), W_{\nu {s'}}(\Omega',\vec{K}'_\perp) )_\nu = e^{ i( \Omega - \Omega') T} \nonumber \\
 & \mathfrak{W}^\dagger_{\nu s} (\Omega,\vec{K}_\perp) \mathfrak{W}_{\nu s'} (\Omega',\vec{K}_\perp')  \sum_{\sigma = \pm} \int_{\mathbb{R}^3} d^3X  e^{s_\nu aZ} \nonumber \\
 & \times e^{i (\vec{K}'_\perp-\vec{K}_\perp) \cdot \vec{X}_\perp}  \mathfrak{K}^*(\sigma s_\nu \Omega,\vec{K}_\perp,s_\nu Z)  \mathfrak{K}(\sigma s_\nu \Omega',\vec{K}_\perp',s_\nu Z) .
\end{align}
Furthermore, one can use the following property for the Bessel function 
\begin{equation}\label{Bessel_conjugate}
K^*_\zeta (\xi) = K_{\zeta^*} (\xi),
\end{equation}
with $\xi \in \mathbb{R}$. A proof for Eq.~(\ref{Bessel_conjugate}) can be obtained by considering the integral representation for the Bessel function [Appendix \ref{Bessel_functions}]. In terms of the functions $\mathfrak{K}(\Omega,\vec{K}_\perp,Z)$, Eq.~(\ref{Bessel_conjugate}) reads [Eq.~(\ref{K_pm})]
\begin{equation}\label{K_conjugate}
\mathfrak{K}^*(\Omega,\vec{K}_\perp,Z) = \mathfrak{K}(- \Omega,\vec{K}_\perp,Z),
\end{equation}
which can be plugged in Eq.~(\ref{W_R_W_R_product_2}) to give
\begin{align} \label{W_R_W_R_product_3}
& ( W_{\nu s}(\Omega,\vec{K}_\perp), W_{\nu {s'}}(\Omega',\vec{K}'_\perp) )_\nu = e^{ i( \Omega - \Omega') T} \nonumber \\
 & \times  \mathfrak{W}^\dagger_{\nu s} (\Omega,\vec{K}_\perp) \mathfrak{W}_{\nu s'} (\Omega',\vec{K}_\perp')  \sum_{\sigma = \pm} \int_{\mathbb{R}^3} d^3X  e^{s_\nu aZ} \nonumber \\
 & \times e^{i (\vec{K}'_\perp-\vec{K}_\perp) \cdot \vec{X}_\perp}  \mathfrak{K}(-\sigma s_\nu \Omega,\vec{K}_\perp,s_\nu Z)  \nonumber \\
 & \times \mathfrak{K}(\sigma s_\nu \Omega',\vec{K}_\perp',s_\nu Z) .
\end{align}

By computing the integral with respect to $X$ and $Y$ in Eq.~(\ref{W_R_W_R_product_3}), one obtains
\begin{align} \label{W_R_W_R_product_4}
& ( W_{\nu s}(\Omega,\vec{K}_\perp), W_{\nu {s'}}(\Omega',\vec{K}'_\perp) )_\nu = 4 \pi^2  \delta^2(\vec{K}_\perp-\vec{K}_\perp') \nonumber \\
 & \times  e^{ i( \Omega - \Omega') T}  \mathfrak{W}^\dagger_{\nu s} (\Omega,\vec{K}_\perp) \mathfrak{W}_{\nu s'} (\Omega',\vec{K}_\perp')  \sum_{\sigma = \pm} \int_{\mathbb{R}} dZ e^{s_\nu aZ} \nonumber \\
 & \times  \mathfrak{K}(-\sigma s_\nu \Omega,\vec{K}_\perp,s_\nu Z)  \mathfrak{K}(\sigma s_\nu \Omega',\vec{K}_\perp,s_\nu Z).
\end{align}
The integral with respect to $Z$, instead, can be computed by using the following identity for Bessel functions
\begin{align}\label{Bessel_orthonormal}
& \int_0^\infty d\xi \left[ K_{-i \zeta-1/2}(\xi) K_{i \zeta'-1/2}(\xi) \right. \nonumber \\
& \left. + K_{i \zeta-1/2}(\xi) K_{-i \zeta'-1/2}(\xi) \right] = \frac{\pi^2 \delta(\zeta-\zeta')}{\cosh(\pi \zeta)}.
\end{align}
A proof for Eq.~(\ref{Bessel_orthonormal}) can be found in Appendix \ref{Bessel_functions}. By replacing $\xi$, $\zeta$ and $\zeta'$ with $\kappa (\vec{K}_\perp) e^{s_\nu aZ} /a$,  $ s_\nu \Omega/(ca)$ and $ s_\nu \Omega'/(ca)$, respectively, in Eq.~(\ref{Bessel_orthonormal}), one obtains the following identity
\begin{align}\label{K_p_orthogonal}
& \sum_{\sigma = \pm} \int_{\mathbb{R}} dZ e^{s_\nu aZ}  \mathfrak{K}(- \sigma s_\nu \Omega,\vec{K}_\perp,s_\nu Z) \mathfrak{K}(\sigma s_\nu \Omega',\vec{K}_\perp,s_\nu Z) \nonumber \\
= & \frac{\pi^2 c a \delta(\Omega-\Omega')}{\kappa (\vec{K}_\perp)}  \left[ \cosh \left( \frac{\beta}{2} \Omega \right) \right]^{-1},
\end{align}
with
\begin{equation}\label{beta}
\beta = \frac{2 \pi}{ca}.
\end{equation}
Equation (\ref{K_p_orthogonal}) can be plugged in Eq.~(\ref{W_R_W_R_product_4}) to give
\begin{align} \label{W_R_W_R_product_5}
& ( W_{\nu s}(\Omega,\vec{K}_\perp), W_{\nu {s'}}(\Omega',\vec{K}'_\perp) )_\nu = \delta(\Omega-\Omega')  \delta^2(\vec{K}_\perp-\vec{K}_\perp') \nonumber \\
 & \times \frac{4 \pi^4  c a  }{\kappa (\vec{K}_\perp) } \mathfrak{W}^\dagger_{\nu s} (\Omega,\vec{K}_\perp) \mathfrak{W}_{\nu s'} (\Omega',\vec{K}_\perp')   \left[ \cosh \left( \frac{\beta}{2} \Omega \right) \right]^{-1},
\end{align}
which is equivalent to Eq.~(\ref{D_scalar_product_orthonormality_W}) only when the following condition is met
\begin{equation}\label{W_orthogonormal}
  \mathfrak{W}^\dagger_{\nu s} (\Omega,\vec{K}_\perp) \mathfrak{W}_{\nu s'} (\Omega',\vec{K}_\perp')  = \delta_{ss'} \frac{\kappa (\vec{K}_\perp)}{4 \pi^4 c a} \cosh \left( \frac{\beta}{2} \Omega \right).
\end{equation}

Equation (\ref{W_orthogonormal}) suggests the definition of the spinor function $\tilde{\mathfrak{W}}_{\nu s}(\Omega, \vec{K}_\perp)$ such that
\begin{equation}\label{W_W_tilde}
\mathfrak{W}_{\nu s}(\Omega,\vec{K}_\perp) = \frac{1}{2 \pi^2} \sqrt{\frac{\kappa (\vec{K}_\perp)}{c a} \cosh \left( \frac{\beta}{2} \Omega \right)} \tilde{\mathfrak{W}}_{\nu s}(\Omega, \vec{K}_\perp).
\end{equation}
The equations defining $\tilde{\mathfrak{W}}_{\nu s}(\Omega, \vec{K}_\perp)$ are given by Eqs.~(\ref{W_p_eigenvalues}) and (\ref{W_orthogonormal}) and explicitly read
\begin{subequations} \label{W_tilde_constraints}
\begin{align}
& c\gamma^0 \gamma^3 \tilde{\mathfrak{W}}_{\nu s}(\Omega, \vec{K}_\perp) =  \tilde{\mathfrak{W}}_{\nu s}(\Omega, \vec{K}_\perp),\label{W_tilde_constraints_b} \\
&  \tilde{\mathfrak{W}}^\dagger_{\nu s} (\Omega,\vec{K}_\perp)  \tilde{\mathfrak{W}}_{\nu s'} (\Omega,\vec{K}_\perp) = \delta_{ss'} \label{W_tilde_constraints_c}.
\end{align}
\end{subequations}
For fixed $\nu$, $\Omega$ and $\vec{K}_\perp$ and for varying $s=\{ 1, 2 \}$, the spinors $\tilde{\mathfrak{W}}_{\nu s}(\Omega, \vec{K}_\perp)$ are an orthonormal basis for the eigenspace of $c \gamma^0 \gamma^3$ with eigenvalue $1$. Hence, the only freedom left by Eq.~(\ref{W_tilde_constraints}) is about the arbitrary choice for the spin basis $\tilde{\mathfrak{W}}_{\nu s}(\Omega, \vec{K}_\perp)$. 

Any change of basis $\tilde{\mathfrak{W}}_{\nu s}(\Omega, \vec{K}_\perp) \mapsto \bar{\tilde{\mathfrak{W}}}_{\nu s'}(\Omega, \vec{K}_\perp)$ is defined by an unitary matrix $\bar{M}_{\nu ss'} (\Omega, \vec{K}_\perp)$ (with matrix indexes $s$ and $s'$), as follows
\begin{subequations}
\begin{align}
& \bar{\tilde{\mathfrak{W}}}_{\nu s}(\Omega, \vec{K}_\perp) = \sum_{s'=1}^2 \bar{M}_{\nu ss'} (\Omega, \vec{K}_\perp) \tilde{\mathfrak{W}}_{\nu s'}(\Omega, \vec{K}_\perp),\\
& \bar{M}_{\nu ss'} (\Omega, \vec{K}_\perp) = \tilde{\mathfrak{W}}^\dagger_{\nu s'}(\Omega, \vec{K}_\perp) \bar{\tilde{\mathfrak{W}}}_{\nu s}(\Omega, \vec{K}_\perp).
\end{align}
\end{subequations}

Notice that for any basis $\tilde{\mathfrak{W}}_{\nu s}(\Omega, \vec{K}_\perp)$ satisfying Eq.~(\ref{W_tilde_constraints}), also the spinor functions $\tilde{\mathfrak{W}}_{\bar{\nu} s}(-\Omega, \vec{K}_\perp)$ (with $\bar{\nu}$ as the opposite of $\nu$, i.e., $\bar{\nu}=\text{L} $ if $\nu= \text{R}$ and $\bar{\nu}= \text{R}$ if $\nu= \text{L}$) satisfy Eq.~(\ref{W_tilde_constraints}). By acknowledging this symmetry, we prove the existence of the change of basis $M_{\nu ss'}(\Omega, \vec{K}_\perp)$ such that
\begin{subequations}
\begin{align}
& \tilde{\mathfrak{W}}_{\bar{\nu} s}(-\Omega, \vec{K}_\perp) = \sum_{s'=1}^2 M_{\nu ss'}(\Omega, \vec{K}_\perp) \tilde{\mathfrak{W}}_{\nu s'}(\Omega, \vec{K}_\perp),\label{W_L_W_R}\\
& M_{\nu ss'} (\Omega, \vec{K}_\perp) =  \tilde{\mathfrak{W}}^\dagger_{\nu s'}(\Omega, \vec{K}_\perp)  \tilde{\mathfrak{W}}_{\bar{\nu} s}(-\Omega, \vec{K}_\perp).\label{M}
\end{align}
\end{subequations}
The unitarity of $M_{\nu ss'} (\Omega, \vec{K}_\perp)$ reads
\begin{subequations}\label{M_unitary}
\begin{align}
& \sum_{s''=1}^2 M^*_{\nu s''s} (\Omega, \vec{K}_\perp)  M_{\nu s''s'} (\Omega, \vec{K}_\perp) = \delta_{ss'}, \\
& \sum_{s''=1}^2 M^*_{\nu ss''} (\Omega, \vec{K}_\perp)  M_{\nu s's''} (\Omega, \vec{K}_\perp) = \delta_{ss'}.
\end{align}
\end{subequations}

Hereafter we do not specify any particular solution of Eq.~(\ref{W_tilde_constraints}). Instead, we consider a general basis $\tilde{\mathfrak{W}}_{\nu s}(\Omega, \vec{K}_\perp)$ for the eigenspace of $c\gamma^0 \gamma^3$ with eigenvalue $1$. We will show that for different choices of $\tilde{\mathfrak{W}}_{\nu s}(\Omega, \vec{K}_\perp)$, different Rindler-Fock representations of the Minkowski vacuum exist. Then, by tracing the left wedge, the dependency of $\tilde{\mathfrak{W}}_{\nu s}(\Omega, \vec{K}_\perp)$ will disappear. Only in Sec.~\ref{Basis_choice}, we will discuss different choices for the spin basis $\tilde{\mathfrak{W}}_{\nu s}(\Omega, \vec{K}_\perp)$.

\section{Bogoliubov transformation}\label{Bogoliubov_transformation}

In the previous sections we considered the Minkowski $(t,\vec{x})$ and the Rindler $(T,\vec{X})$ spacetimes and we studied the respective Dirac fields $\hat{\psi}(t,\vec{x})$ and $\hat{\Psi}_\nu (T,\vec{X})$. We defined the operators $\hat{c}_s(\vec{k})$, $\hat{d}_s(\vec{k})$, $\hat{C}_{\nu s}(\Omega,\vec{K}_\perp)$ and $\hat{D}_{\nu s}(\Omega,\vec{K}_\perp)$ as the annihilators of positive and negative frequency modes for each spacetime.

In this section, we consider both the Minkowski $(t,\vec{x})$ and the Rindler $(T,\vec{X})$ spacetimes to describe the inertial and the accelerated frame of a flat spacetime. The operators $\hat{\psi}(t,\vec{x})$ and $\hat{\Psi}_\nu (T,\vec{X})$ define the same Dirac field in each coordinate system. We compute the Bogoliubov transformation relating Minkowski ($\hat{c}_s(\vec{k})$ and $\hat{d}_s(\vec{k})$) and Rindler ($\hat{C}_{\nu s}(\Omega,\vec{K}_\perp)$ and $\hat{D}_{\nu s}(\Omega,\vec{K}_\perp)$) operators. We follow the same method presented in \cite{falcone2022} for scalar fields. A different approach can instead be found in \cite{Oriti}.

Equation (\ref{D_scalar_product_orthonormality_u_v}) can be used to invert Eq.~(\ref{free_Dirac_field}) as follows
\begin{align}\label{Bogoliubov_transformations_1_Rindler}
& \hat{c}_s(\vec{k}) = ( u_s(\vec{k}), \hat{\psi})_\text{M}, & \hat{d}_s^\dagger(\vec{k}) = (v_s(\vec{k}),  \hat{\psi})_\text{M}.
\end{align}
Equation (\ref{Bogoliubov_transformations_1_Rindler}) explicitly reads
\begin{subequations}\label{Bogoliubov_transformations_2_Rindler}
\begin{align}
\hat{c}_s(\vec{k}) = & \int_{\mathbb{R}^3} d^3x u^\dagger_s(\vec{k},t,\vec{x}) \hat{\psi}(t,\vec{x}), \\
\hat{d}_s^\dagger(\vec{k}) = & \int_{\mathbb{R}^3} d^3x v^\dagger_s(\vec{k},t,\vec{x}) \hat{\psi}(t,\vec{x}) .
\end{align}
\end{subequations}

The Dirac field transforms as a spinor field under diffeomorphisms. In the case of Rindler coordinates, the transformation $\hat{\Psi}_\nu \mapsto \hat{\psi}$ reads \cite{Oriti}
\begin{align}\label{scalar_transformation_Rindler_inverse}
\hat{\psi}(t,\vec{x}) = & \sum_{\nu=\{\text{L},\text{R}\}} \theta(s_\nu z) \exp \left( \frac{1}{2} \gamma^0 \gamma^3 T_\nu(t,\vec{x}) \right) \nonumber \\
& \times  \hat{\Psi}_\nu(T_\nu(t,\vec{x}),\vec{X}_\nu(t,\vec{x})),
\end{align}
where the functions $T_\nu(t,\vec{x}) $ and $\vec{X}_\nu(t,\vec{x})$ map the Minkowski coordinates $(t,\vec{x})$ to the Rindler coordinates $(T,\vec{X})$ and are the inverse of Eq.~(\ref{Rindler_coordinates_transformation}). When $t=0$, the transformation (\ref{scalar_transformation_Rindler_inverse}) reads
\begin{equation}\label{scalar_transformation_Rindler_inverse_2}
\hat{\psi}(0,\vec{x}) = \sum_{\nu=\{\text{L},\text{R}\}} \theta(s_\nu z) \hat{\Psi}_\nu(0,\vec{X}_\nu(\vec{x})),
\end{equation}
where $\vec{X}_\nu(\vec{x})$ is the coordinate transformation from the Minkowski to the $\nu$-Rindler spacetime when $t  = 0$. The function $\vec{X}_\nu(\vec{x})$ explicitly reads
\begin{equation}
\vec{X}_\nu(\vec{x}_\perp,z) = (\vec{x}_\perp, Z_\nu(z)),
\end{equation}
where $Z_\nu(z)$ is such that
\begin{equation}\label{z_Z}
a z = s_\nu \exp(s_\nu a Z_\nu(z)),
\end{equation}
for any $z$ such that $s_\nu z>0$. By choosing $t=0$ and using Eq.~(\ref{scalar_transformation_Rindler_inverse_2}) in Eq.~(\ref{Bogoliubov_transformations_2_Rindler}) one obtains
\begin{subequations}\label{Bogoliubov_transformations_3_Rindler}
\begin{align}
\hat{c}_s(\vec{k}) = & \sum_{\nu=\{\text{L},\text{R}\}}  \int_{\mathbb{R}^3} d^3x \theta(s_\nu z) u^\dagger_s(\vec{k},0,\vec{x}) \hat{\Psi}_\nu(0,\vec{X}_\nu(\vec{x})),\label{Bogoliubov_transformations_3_Rindler_a} \\
\hat{d}_s^\dagger(\vec{k}) = & \sum_{\nu=\{\text{L},\text{R}\}}  \int_{\mathbb{R}^3} d^3x \theta(s_\nu z) v^\dagger_s(\vec{k},0,\vec{x}) \hat{\Psi}_\nu(0,\vec{X}_\nu(\vec{x}))\label{Bogoliubov_transformations_3_Rindler_b}.
\end{align}
\end{subequations}

By plugging Eq.~(\ref{free_Dirac_field_Rindler_2}) in Eq.~(\ref{Bogoliubov_transformations_3_Rindler}) one is able to related the Minkowski operators $\hat{c}_s(\vec{k})$ and $\hat{d}_s(\vec{k})$ to the Rindler operators $\hat{C}_{\nu s}(\Omega,\vec{K}_\perp)$ and $\hat{D}_{\nu s}(\Omega,\vec{K}_\perp)$ through the following Bogoliubov transformation
\begin{subequations}\label{Bogoliubov_transformations_3_Rindler_2}
\begin{align}
& \hat{c}_s(\vec{k}) = \sum_{\nu=\{\text{L},\text{R}\}}  \sum_{s'=1}^2 \int_\mathbb{R} d\Omega \int_{\mathbb{R}^2} d^2 K_\perp  \nonumber \\
& \times  \int_{\mathbb{R}^3} d^3x \theta(s_\nu z) u^\dagger_s(\vec{k},0,\vec{x}) W_{\nu s'}(\Omega,\vec{K}_\perp,0,\vec{X}_\nu(\vec{x}))  \nonumber \\
& \times \left[ \theta(\Omega) \hat{C}_{\nu s'}(\Omega,\vec{K}_\perp)+  \theta(-\Omega)\hat{D}_{\nu s'}^\dagger(-\Omega,-\vec{K}_\perp) \right], \\
& \hat{d}_s^\dagger(\vec{k}) = \sum_{\nu=\{\text{L},\text{R}\}}  \sum_{s'=1}^2 \int_\mathbb{R} d\Omega \int_{\mathbb{R}^2} d^2 K_\perp    \nonumber \\
& \times  \int_{\mathbb{R}^3} d^3x \theta(s_\nu z) v^\dagger_s(\vec{k},0,\vec{x}) W_{\nu s'}(-\Omega,-\vec{K}_\perp,0,\vec{X}_\nu(\vec{x}))\nonumber \\
& \times \left[ \theta(-\Omega) \hat{C}_{\nu s'}(-\Omega,-\vec{K}_\perp)+  \theta(\Omega)\hat{D}_{\nu s'}^\dagger(\Omega,\vec{K}_\perp) \right].
\end{align}
\end{subequations}
By using Eqs.~(\ref{free_Dirac_field_modes}) and (\ref{UVW_UVW_tilde}) and by performing the integration with respect to $x$ and $y$, the Bogoliubov transformation (\ref{Bogoliubov_transformations_3_Rindler_2}) reads
\begin{subequations}\label{Bogoliubov_transformations_3_Rindler_3}
\begin{align}
& \hat{c}_s(\vec{k}) = \sqrt{2\pi} \sum_{\nu=\{\text{L},\text{R}\}}  \sum_{s'=1}^2 \int_\mathbb{R} d\Omega \int_{\mathbb{R}^2} d^2 K_\perp  \delta^2(\vec{k}_\perp - \vec{K}_\perp)  \nonumber \\
& \times  \int_{\mathbb{R}} dz \theta(s_\nu z) e^{ - ik_3 z } \tilde{u}^\dagger_s(\vec{k}) \tilde{W}_{\nu s'}(\Omega,\vec{k}_\perp,Z_\nu(z))\nonumber \\
& \times \left[ \theta(\Omega) \hat{C}_{\nu s'}(\Omega,\vec{K}_\perp)+  \theta(-\Omega)\hat{D}_{\nu s'}^\dagger(-\Omega,-\vec{K}_\perp) \right] , \label{Bogoliubov_transformations_3_Rindler_3_a}\\
& \hat{d}_s^\dagger(\vec{k}) = \sqrt{2\pi} \sum_{\nu=\{\text{L},\text{R}\}}  \sum_{s'=1}^2 \int_\mathbb{R} d\Omega \int_{\mathbb{R}^2} d^2 K_\perp  \delta^2(\vec{k}_\perp - \vec{K}_\perp)  \nonumber \\
& \times  \int_{\mathbb{R}} dz \theta(s_\nu z)e^{ ik_3 z } \tilde{v}^\dagger_s(\vec{k}) \tilde{W}_{\nu s'}(-\Omega,-\vec{k}_\perp,Z_\nu(z)) \nonumber \\
& \times \left[ \theta(-\Omega) \hat{C}_{\nu s'}(-\Omega,-\vec{K}_\perp)+  \theta(\Omega)\hat{D}_{\nu s'}^\dagger(\Omega,\vec{K}_\perp) \right]. \label{Bogoliubov_transformations_3_Rindler_3_b}
\end{align}
\end{subequations}

We now focus on the scalar products $\tilde{u}^\dagger_s(\vec{k}) \tilde{W}_{\nu s'}(\Omega,\vec{k}_\perp,Z_\nu(z))$ and $\tilde{v}^\dagger_s(\vec{k}) \tilde{W}_{\nu s'}(-\Omega,-\vec{k}_\perp,Z_\nu(z))$ that appear in Eqs.~(\ref{Bogoliubov_transformations_3_Rindler_3_a}) and (\ref{Bogoliubov_transformations_3_Rindler_3_b}), respectively. By using Eqs.~(\ref{U_U_p_U_m}), (\ref{U_pm_s_W_pm_s}) and (\ref{W_pm_W}), one can write
\begin{align}\label{W_tilde_W_tilde}
& \tilde{W}_{\nu s'}(\Omega,\vec{K}_\perp,Z_\nu(z)) = \sum_{\sigma=\pm}  \mathfrak{K} (\sigma s_\nu \Omega,\vec{K}_\perp,s_\nu Z) \nonumber \\
 &\times  \left[ \mathfrak{G}_\nu(\vec{K}_\perp) \right]^{(1 - \sigma)/2} \mathfrak{W}_{\nu s}(\Omega,\vec{K}_\perp).
\end{align}
Hence, to obtain $\tilde{u}^\dagger_s(\vec{k}) \tilde{W}_{\nu s'}(\Omega,\vec{k}_\perp,Z_\nu(z))$ and $\tilde{v}^\dagger_s(\vec{k}) \tilde{W}_{\nu s'}(-\Omega,-\vec{k}_\perp,Z_\nu(z))$, one firstly has to compute the following scalar products: $\tilde{u}^\dagger_s(\vec{k}) \mathfrak{G}_\nu(\vec{k}_\perp)\mathfrak{W}_{\nu s'}(\Omega,\vec{k}_\perp)$ and $\tilde{v}^\dagger_s(\vec{k}) \mathfrak{G}_\nu(-\vec{k}_\perp)\mathfrak{W}_{\nu s'}(-\Omega,-\vec{k}_\perp)$. The former can be obtained by using Eqs.~(\ref{gamma_matrices_identities}), (\ref{Dirac_u_tilde}), (\ref{Gamma}), (\ref{W_p_eigenvalues}) and (\ref{G_antihermitian})
\begin{align} \label{u_tilde_Gamma_W_tilde}
& \tilde{u}^\dagger_s(\vec{k}) \mathfrak{G}_\nu(\vec{k}_\perp)\mathfrak{W}_{\nu s'}(\Omega,\vec{k}_\perp) \nonumber \\
= & - \left[ \mathfrak{G}_\nu(\vec{k}_\perp) \tilde{u}_s(\vec{k}) \right]^\dagger \mathfrak{W}_{\nu s'}(\Omega,\vec{k}_\perp) \nonumber \\
 = &\frac{- s_\nu i c} {\kappa (\vec{k}_\perp)} \left[ \gamma^0 \left( k_1 \gamma^1 + k_2 \gamma^2 + \frac{m c}{\hbar} \right) \tilde{u}_s(\vec{k}) \right]^\dagger \mathfrak{W}_{\nu s'}(\Omega, \vec{k}_\perp) \nonumber \\
 = &\frac{-s_\nu i c} {\kappa (\vec{k}_\perp)} \left\lbrace \gamma^0 \left[ \omega(\vec{k}) \gamma^0 - k_3 \gamma^3 \right] \tilde{u}_s(\vec{k}) \right\rbrace^\dagger \mathfrak{W}_{\nu s'}(\Omega, \vec{k}_\perp)\nonumber \\
 = &\frac{-s_\nu i} {\kappa (\vec{k}_\perp)} \left\lbrace \left[ \frac{\omega(\vec{k})}{c} -  c k_3 \gamma^0 \gamma^3 \right] \tilde{u}_s(\vec{k}) \right\rbrace^\dagger \mathfrak{W}_{\nu s'}(\Omega, \vec{k}_\perp)\nonumber \\
 = &\frac{-s_\nu i} {\kappa (\vec{k}_\perp)}  \tilde{u}^\dagger_s(\vec{k}) \left[ \frac{\omega(\vec{k})}{c} - c k_3 \gamma^0 \gamma^3  \right] \mathfrak{W}_{\nu s'}(\Omega, \vec{k}_\perp) \nonumber \\
 = &\frac{-s_\nu i} {\kappa (\vec{k}_\perp)}  \left[ \frac{\omega(\vec{k})}{c} - k_3 \right] \tilde{u}^\dagger_s(\vec{k}) \mathfrak{W}_{\nu s'}(\Omega, \vec{k}_\perp).
\end{align}
Similarly, for the second scalar product one can use Eq.~(\ref{Dirac_v_tilde}) instead of Eq.~(\ref{Dirac_u_tilde})
\begin{align} \label{v_tilde_Gamma_W_tilde}
& \tilde{v}^\dagger_s(\vec{k}) \mathfrak{G}_\nu(-\vec{k}_\perp)\mathfrak{W}_{\nu s'}(-\Omega,-\vec{k}_\perp) \nonumber \\
= & - \left[ \mathfrak{G}_\nu(-\vec{k}_\perp) \tilde{v}_s(\vec{k}) \right]^\dagger \mathfrak{W}_{\nu s'}(-\Omega,-\vec{k}_\perp) \nonumber \\
 = &\frac{s_\nu i c} {\kappa (\vec{k}_\perp)} \left[ \gamma^0 \left( k_1 \gamma^1 + k_2 \gamma^2 - \frac{m c}{\hbar} \right) \tilde{v}_s(\vec{k}) \right]^\dagger  \nonumber \\
& \times \mathfrak{W}_{\nu s'}(-\Omega, -\vec{k}_\perp) \nonumber \\
 = &\frac{s_\nu i c} {\kappa (\vec{k}_\perp)} \left\lbrace \gamma^0 \left[ \omega(\vec{k}) \gamma^0 - k_3 \gamma^3 \right] \tilde{v}_s(\vec{k}) \right\rbrace^\dagger \mathfrak{W}_{\nu s'}(-\Omega, -\vec{k}_\perp)\nonumber \\
 = &\frac{s_\nu i} {\kappa (\vec{k}_\perp)} \left\lbrace \left[ \frac{\omega(\vec{k})}{c} -  c k_3 \gamma^0 \gamma^3 \right] \tilde{v}_s(\vec{k}) \right\rbrace^\dagger \mathfrak{W}_{\nu s'}(-\Omega, -\vec{k}_\perp)\nonumber \\
 = &\frac{s_\nu i} {\kappa (\vec{k}_\perp)}  \tilde{v}^\dagger_s(\vec{k}) \left[ \frac{\omega(\vec{k})}{c} - c k_3 \gamma^0 \gamma^3  \right] \mathfrak{W}_{\nu s'}(-\Omega, -\vec{k}_\perp) \nonumber \\
 = &\frac{s_\nu i} {\kappa (\vec{k}_\perp)}  \left[ \frac{\omega(\vec{k})}{c} - k_3 \right] \tilde{v}^\dagger_s(\vec{k}) \mathfrak{W}_{\nu s'}(-\Omega, -\vec{k}_\perp).
\end{align}

Equations (\ref{omega_k}) and (\ref{kappa_k_perp}) lead to
\begin{align}
& \frac{\omega^2(\vec{k})}{c^2}  - k_3^2 = \kappa^2 (\vec{k}_\perp), && \omega (\vec{k})> 0, && \kappa (\vec{k}_\perp) > 0,
\end{align}
which suggests the definition of the function $\vartheta(\vec{k})$ such that
\begin{subequations}\label{theta_k}
\begin{align}
& \omega (\vec{k}) = c \kappa (\vec{k}_\perp) \cosh(\vartheta(\vec{k})),\label{omega_theta} \\
& k_3 = \kappa (\vec{k}_\perp) \sinh(\vartheta(\vec{k})) \label{k_3_theta}.
\end{align}
\end{subequations}
In this way, Eqs.~(\ref{u_tilde_Gamma_W_tilde}) and (\ref{v_tilde_Gamma_W_tilde}) read
\begin{subequations}\label{uv_tilde_Gamma_W_tilde_2}
\begin{align}
& \tilde{u}^\dagger_s(\vec{k}) \mathfrak{G}_\nu(\vec{k}_\perp)\mathfrak{W}_{\nu s'}(\Omega,\vec{k}_\perp) \nonumber \\
 = & \exp \left(- s_\nu i \frac{\pi}{2} -\vartheta(\vec{k}) \right)   \tilde{u}^\dagger_s(\vec{k})  \mathfrak{W}_{\nu s'}(\Omega, \vec{k}_\perp), \\
& \tilde{v}^\dagger_s(\vec{k}) \mathfrak{G}_\nu(-\vec{k}_\perp)\mathfrak{W}_{\nu s'}(-\Omega,-\vec{k}_\perp) \nonumber \\
 = & \exp \left(s_\nu i \frac{\pi}{2} -\vartheta(\vec{k}) \right)  \tilde{v}^\dagger_s(\vec{k}) \mathfrak{W}_{\nu s'}(-\Omega, -\vec{k}_\perp),
\end{align}
\end{subequations}
which means that
\begin{subequations}\label{uv_tilde_Gamma_W_tilde_3}
\begin{align}
& \tilde{u}^\dagger_s(\vec{k}) \left[ \mathfrak{G}_\nu(\vec{k}_\perp) \right]^{(1 - \sigma)/2} \mathfrak{W}_{\nu s'}(\Omega,\vec{k}_\perp) \nonumber \\
 = & \exp \left( \frac{\sigma-1}{2} \left[s_\nu i \frac{\pi}{2} +\vartheta(\vec{k})\right] \right)   \tilde{u}^\dagger_s(\vec{k})  \mathfrak{W}_{\nu s'}(\Omega, \vec{k}_\perp), \\
& \tilde{v}^\dagger_s(\vec{k})\left[ \mathfrak{G}_\nu(-\vec{k}_\perp) \right]^{(1 - \sigma)/2} \mathfrak{W}_{\nu s'}(-\Omega,-\vec{k}_\perp) \nonumber \\
 = & \exp \left(\frac{\sigma-1}{2} \left[-s_\nu i \frac{\pi}{2} +\vartheta(\vec{k})\right]  \right)   \tilde{v}^\dagger_s(\vec{k}) \mathfrak{W}_{\nu s'}(-\Omega, -\vec{k}_\perp).
\end{align}
\end{subequations}
Equations (\ref{K_conjugate}), (\ref{W_tilde_W_tilde}) and (\ref{uv_tilde_Gamma_W_tilde_3}) and the fact that $\kappa (\vec{K}_\perp)$ and $ \mathfrak{K} (\Omega,\vec{K}_\perp,Z)$ are even with respect to $\vec{K}_\perp$ [Eqs.~(\ref{kappa_k_perp}) and (\ref{K_pm})] allow to compute the following scalar products
\begin{subequations}\label{uv_tilde_Gamma_W_tilde_4}
\begin{align}
& \tilde{u}^\dagger_s(\vec{k}) \left[ \mathfrak{G}_\nu(\vec{k}_\perp) \right]^{(1 - \sigma)/2} \tilde{W}_{\nu s'}(\Omega,\vec{k}_\perp,Z_\nu(z)) \nonumber \\
 = & \sum_{\sigma=\pm}  \mathfrak{K} (\sigma s_\nu \Omega,\vec{k}_\perp,Z_\nu(z)) \exp \left( \frac{\sigma-1}{2} \left[s_\nu i \frac{\pi}{2} +\vartheta(\vec{k})\right] \right)  \nonumber \\
 &\times   \tilde{u}^\dagger_s(\vec{k})  \mathfrak{W}_{\nu s'}(\Omega, \vec{k}_\perp), \\
& \tilde{v}^\dagger_s(\vec{k})\left[ \mathfrak{G}_\nu(-\vec{k}_\perp) \right]^{(1 - \sigma)/2} \tilde{W}_{\nu s'}(-\Omega,-\vec{k}_\perp,Z_\nu(z))\nonumber \\
 = & \sum_{\sigma=\pm} \mathfrak{K}^* (\sigma s_\nu \Omega,\vec{k}_\perp,Z_\nu(z))  \nonumber \\
 &\times \exp \left(\frac{\sigma-1}{2} \left[-s_\nu i \frac{\pi}{2} +\vartheta(\vec{k})\right]  \right)  \tilde{v}^\dagger_s(\vec{k}) \mathfrak{W}_{\nu s'}(-\Omega, -\vec{k}_\perp).
\end{align}
\end{subequations}

By plugging Eq.~(\ref{uv_tilde_Gamma_W_tilde_4}) in Eq.~(\ref{Bogoliubov_transformations_3_Rindler_3}) and using Eq.~(\ref{W_W_tilde}), one obtains
\begin{subequations}\label{Bogoliubov_transformations_3_Rindler_4}
\begin{align}
& \hat{c}_s(\vec{k}) = \sum_{\nu=\{\text{L},\text{R}\}}  \sum_{s'=1}^2 \int_\mathbb{R} d\Omega \int_{\mathbb{R}^2} d^2 K_\perp   \nonumber \\
& \times  \alpha_{\nu}(\vec{k},\Omega,\vec{K}_\perp) \tilde{u}^\dagger_s(\vec{k})  \tilde{\mathfrak{W}}_{\nu s'}(\Omega, \vec{K}_\perp) \nonumber \\
& \times \left[ \theta(\Omega) \hat{C}_{\nu s'}(\Omega,\vec{K}_\perp)+  \theta(-\Omega)\hat{D}_{\nu s'}^\dagger(-\Omega,-\vec{K}_\perp) \right] ,\label{Bogoliubov_transformations_3_Rindler_4_a}\\
& \hat{d}_s^\dagger(\vec{k}) = \sum_{\nu=\{\text{L},\text{R}\}}  \sum_{s'=1}^2 \int_\mathbb{R} d\Omega \int_{\mathbb{R}^2} d^2 K_\perp   \nonumber \\
& \times  \alpha^*_{\nu}(\vec{k},\Omega,\vec{K}_\perp) \tilde{v}^\dagger_s(\vec{k})  \tilde{\mathfrak{W}}_{\nu s'}(-\Omega, -\vec{K}_\perp) \nonumber \\
& \times \left[ \theta(-\Omega) \hat{C}_{\nu s'}(-\Omega,-\vec{K}_\perp)+  \theta(\Omega)\hat{D}_{\nu s'}^\dagger(\Omega,\vec{K}_\perp) \right],\label{Bogoliubov_transformations_3_Rindler_4_b}
\end{align}
\end{subequations}
with the following Bogoliubov coefficient
\begin{align}\label{Bogoliubov_coefficient}
& \alpha_{\nu}(\vec{k},\Omega,\vec{K}_\perp) = \frac{1}{\pi} \delta^2(\vec{k}_\perp - \vec{K}_\perp) \sqrt{\frac{\kappa (\vec{k}_\perp)}{2 \pi c a} \cosh \left( \frac{\beta}{2} \Omega \right)}  \nonumber \\
& \times \exp \left( -s_\nu i \frac{\pi}{4} - \frac{\vartheta(\vec{k})}{2} \right) I_\nu (\vec{k},\Omega)
\end{align}
and with
\begin{subequations}
\begin{align}
& I_\nu (\vec{k},\Omega) = \sum_{\sigma = \pm} \tilde{I}_\nu (\vec{k},\sigma \Omega) \exp \left( \sigma s_\nu i \frac{\pi}{4} + \sigma \frac{\vartheta(\vec{k})}{2} \right), \label{I_I_tilde} \\
&  \tilde{I}_\nu (\vec{k},\Omega) =  \int_{\mathbb{R}} dz \theta(s_\nu z) e^{ - ik_3 z }  \mathfrak{K} (s_\nu \Omega,\vec{k}_\perp, s_\nu Z_\nu(z)). \label{I_tilde}
\end{align}
\end{subequations}

The integral appearing in Eq.~(\ref{I_tilde}) can be computed by considering the following identity for Bessel functions
\begin{equation}\label{Bessel_integral_representation_final}
\int_{\mathbb{R}} d\xi \theta(\xi) e^{- i \xi \sinh(\tau)} K_\zeta (\xi)   =  \frac{\pi \sin \left( \zeta \left( \frac{\pi}{2} - i \tau \right) \right)}{ \sin(\pi \zeta) \cosh(\tau)}.
\end{equation}
A prove for Eq.~(\ref{Bessel_integral_representation_final}) can be found in Appendix \ref{Bessel_functions}. By replacing the variables $\xi$, $\zeta$ and $\tau$ with, respectively, $s_\nu \kappa(\vec{k}_\perp) z$, $ s_\nu i \Omega/(ca) - 1/2$ and $s_\nu \vartheta(\vec{k})$ and by dividing the equation with $\kappa(\vec{k}_\perp)$, one obtains
\begin{align}\label{Bessel_integral_representation_6}
& \int_{\mathbb{R}} dz \theta(s_\nu z) \exp( -  i \kappa(\vec{k}_\perp) \sinh(\vartheta(\vec{k})) z ) \nonumber \\
& \times K_{s_\nu i \Omega / (c a) - 1/2} \left( s_\nu \kappa (\vec{k}_\perp) z \right)  \nonumber \\
 = &  \frac{\pi \sin \left( s_\nu i \frac{\pi \Omega}{2 c a} - \frac{\pi}{4} +  \frac{\vartheta(\vec{k}) \Omega}{c a} + s_\nu i \frac{\vartheta(\vec{k})}{2}  \right)}{\kappa(\vec{k}_\perp) \sin\left( s_\nu i \frac{\pi \Omega}{c a} - \frac{\pi}{2}  \right) \cosh(\vartheta(\vec{k})) }.
\end{align}
By using Eqs.~(\ref{K_pm}), (\ref{beta}), (\ref{z_Z}), (\ref{k_3_theta}), (\ref{I_tilde}) and (\ref{Bessel_integral_representation_6}), one can compute $\tilde{I}_\nu(\vec{k},\Omega)$ in the following way
\begin{align}\label{I_tilde_2}
 \tilde{I}_\nu(\vec{k},\Omega)= & \int_{\mathbb{R}} dz \theta(s_\nu z) \exp( - i \kappa(\vec{k}_\perp) \sinh(\vartheta(\vec{k})) z ) \nonumber \\
 & \times  K_{s_\nu i \Omega / (c a) - 1/2} \left( \kappa (\vec{k}_\perp) \frac{e^{s_\nu a Z_\nu(z)}}{a} \right)  \nonumber \\
= & \int_{\mathbb{R}} dz \theta(s_\nu z) \exp( - i \kappa(\vec{k}_\perp) \sinh(\vartheta(\vec{k})) z ) \nonumber \\
 & \times  K_{s_\nu i \Omega / (c a) - 1/2} \left( s_\nu \kappa (\vec{k}_\perp) z \right)  \nonumber \\
 = &  \frac{\pi \sin \left( s_\nu i \frac{\beta \Omega}{4} - \frac{\pi}{4} +  \frac{\vartheta(\vec{k}) \Omega}{c a} + s_\nu i \frac{\vartheta(\vec{k})}{2}  \right)}{\kappa(\vec{k}_\perp) \sin\left( s_\nu i \frac{\beta \Omega}{2} - \frac{\pi}{2}  \right) \cosh(\vartheta(\vec{k})) } \nonumber \\
= & - \frac{s_\nu \pi \sin \left( i \frac{\beta \Omega}{4} - s_\nu \frac{\pi}{4} + s_\nu \frac{\vartheta(\vec{k}) \Omega}{c a} + i \frac{\vartheta(\vec{k})}{2} \right) }{\kappa(\vec{k}_\perp) \cosh\left( \frac{\beta \Omega}{2}  \right) \cosh(\vartheta(\vec{k})) }.
\end{align}

By plugging Eq.~(\ref{I_tilde_2}) in Eq.~(\ref{I_I_tilde}) one can compute the following function
\begin{align}\label{I}
& I_\nu (\vec{k},\Omega) \nonumber \\
= &- s_\nu \pi \left[ \kappa(\vec{k}_\perp) \cosh\left( \frac{\beta \Omega}{2}  \right) \cosh(\vartheta(\vec{k})) \right]^{-1}\nonumber \\
& \times  \sum_{\sigma = \pm} \exp \left( \sigma s_\nu i \frac{\pi}{4} + \sigma \frac{\vartheta(\vec{k})}{2} \right) \nonumber \\
& \times \sin \left( \sigma i \frac{\beta \Omega}{4} - s_\nu \frac{\pi}{4} + \sigma s_\nu \frac{\vartheta(\vec{k}) \Omega}{c a} + i \frac{\vartheta(\vec{k})}{2} \right) \nonumber \\
= &s_\nu i \pi \left[ 2 \kappa(\vec{k}_\perp)  \cosh\left( \frac{\beta \Omega}{2}  \right) \cosh(\vartheta(\vec{k})) \right]^{-1} \nonumber \\
& \times  \sum_{\sigma = \pm} \left[ \exp \left( s_\nu i \frac{\sigma - 1}{4} \pi  + \frac{\sigma-1}{2} \vartheta(\vec{k}) \right.\right. \nonumber \\
& \left.  - \sigma \frac{\beta \Omega}{4} + \sigma s_\nu i \frac{\vartheta(\vec{k}) \Omega}{c a}\right) - \exp \left( s_\nu i \frac{\sigma + 1}{4} \pi\right.\nonumber \\
& \left.  \left.+ \frac{\sigma+1}{2} \vartheta(\vec{k})  + \sigma  \frac{\beta \Omega}{4}  - \sigma s_\nu i \frac{\vartheta(\vec{k}) \Omega}{c a} \right)  \right]\nonumber \\
= & s_\nu i \pi \left[ 2 \kappa(\vec{k}_\perp) \cosh\left( \frac{\beta \Omega}{2}  \right) \cosh(\vartheta(\vec{k})) \right]^{-1}    \nonumber \\
& \times \left[ - \exp \left(  s_\nu i \frac{\pi}{2} + \vartheta(\vec{k}) +  \frac{\beta \Omega}{4} - s_\nu i \frac{\vartheta(\vec{k}) \Omega}{c a}  \right)  \right. \nonumber \\
& \left. + \exp \left( - s_\nu i \frac{\pi}{2} - \vartheta(\vec{k}) + \frac{\beta \Omega}{4} - s_\nu i  \frac{\vartheta(\vec{k}) \Omega}{c a} \right)\right] \nonumber \\
= &  s_\nu i \pi \left[ 2 \kappa(\vec{k}_\perp)  \cosh\left( \frac{\beta \Omega}{2}  \right) \cosh(\vartheta(\vec{k})) \right]^{-1}    \nonumber \\
 &  \times    \exp \left(  \frac{\beta \Omega}{4} - s_\nu i  \frac{\vartheta(\vec{k}) \Omega}{c a} \right) \left[ - s_\nu i e^{\vartheta(\vec{k})}  - s_\nu i e^{-\vartheta(\vec{k})} \right] \nonumber \\
= &  \pi \left[ \kappa(\vec{k}_\perp)  \cosh\left( \frac{\beta \Omega}{2}  \right) \right]^{-1}   \exp \left(  \frac{\beta \Omega}{4} - s_\nu i  \frac{\vartheta(\vec{k}) \Omega}{c a} \right).
\end{align}

Equation (\ref{I}) can be used in Eq.~(\ref{Bogoliubov_coefficient}) to obtain the final expression for the Bogoliubov coefficients
\begin{align}\label{Bogoliubov_coefficient_2}
& \alpha_{\nu}(\vec{k},\Omega,\vec{K}_\perp) = \delta^2(\vec{k}_\perp - \vec{K}_\perp) \nonumber \\
& \times  \frac{\exp \left( - s_\nu i \frac{\pi}{4} - \frac{\vartheta(\vec{k})}{2} + \frac{\beta \Omega}{4} - s_\nu i  \frac{\vartheta(\vec{k}) \Omega}{c a} \right)}{\sqrt{2 \pi c a \kappa (\vec{k}_\perp) \cosh \left( \frac{\beta}{2} \Omega \right)}}.
\end{align}
By using the fact that $s_{\bar{\nu}}=-s_\nu$, Eq.~(\ref{Bogoliubov_coefficient_2}) leads to the following identity
\begin{align}\label{alpha_beta_Unruh}
 \alpha_{\bar{\nu}}(\vec{k},-\Omega,\vec{K}_\perp) = & s_\nu i e^{-\beta \Omega / 2}  \alpha_{\nu}(\vec{k},\Omega,\vec{K}_\perp),
\end{align} 
which can be used in Eq.~(\ref{Bogoliubov_transformations_3_Rindler_4}) to relate operators of opposite frequency and wedge. By inverting the variables $\Omega \mapsto -\Omega$ and $\nu \mapsto \bar{\nu}$ when  $\Omega < 0$, Eq.~(\ref{Bogoliubov_transformations_3_Rindler_4}) reads
\begin{subequations}\label{Bogoliubov_transformations_3_Rindler_5}
\begin{align}
& \hat{c}_s(\vec{k}) = \sum_{\nu=\{\text{L},\text{R}\}}  \sum_{s'=1}^2 \int_0^\infty d\Omega \int_{\mathbb{R}^2} d^2 K_\perp   \nonumber \\
& \times  \tilde{u}^\dagger_s(\vec{k}) \left[ \alpha_{\nu}(\vec{k},\Omega,\vec{K}_\perp) \tilde{\mathfrak{W}}_{\nu s'}(\Omega, \vec{K}_\perp) \hat{C}_{\nu s'}(\Omega,\vec{K}_\perp) \right. \nonumber \\
&  \left. +  \alpha_{\bar{\nu}}(\vec{k},-\Omega,\vec{K}_\perp) \tilde{\mathfrak{W}}_{\bar{\nu} s'}(-\Omega, \vec{K}_\perp) \hat{D}_{\bar{\nu} s'}^\dagger(\Omega,-\vec{K}_\perp) \right] ,\\
& \hat{d}_s^\dagger(\vec{k}) = \sum_{\nu=\{\text{L},\text{R}\}}  \sum_{s'=1}^2 \int_0^\infty d\Omega \int_{\mathbb{R}^2} d^2 K_\perp   \nonumber \\
& \times   \tilde{v}^\dagger_s(\vec{k}) \left[ \alpha^*_{\bar{\nu}}(\vec{k},-\Omega,\vec{K}_\perp) \tilde{\mathfrak{W}}_{\bar{\nu} s'}(\Omega, -\vec{K}_\perp)  \hat{C}_{\bar{\nu} s'}(\Omega,-\vec{K}_\perp)\right. \nonumber \\
&  \left. +  \alpha^*_{\nu}(\vec{k},\Omega,\vec{K}_\perp) \tilde{\mathfrak{W}}_{\nu s'}(-\Omega, -\vec{K}_\perp) \hat{D}_{\nu s'}^\dagger(\Omega,\vec{K}_\perp) \right].
\end{align}
\end{subequations}
By plugging Eq.~(\ref{alpha_beta_Unruh}) in Eq.~(\ref{Bogoliubov_transformations_3_Rindler_5}),
\begin{subequations}\label{Bogoliubov_transformations_3_Rindler_6}
\begin{align}
& \hat{c}_s(\vec{k}) = \sum_{\nu=\{\text{L},\text{R}\}}  \sum_{s'=1}^2 \int_0^\infty d\Omega \int_{\mathbb{R}^2} d^2 K_\perp   \alpha_{\nu}(\vec{k},\Omega,\vec{K}_\perp) \nonumber \\
& \times  \tilde{u}^\dagger_s(\vec{k}) \left[ \tilde{\mathfrak{W}}_{\nu s'}(\Omega, \vec{K}_\perp) \hat{C}_{\nu s'}(\Omega,\vec{K}_\perp) \right. \nonumber \\
&  \left. +  s_\nu i e^{-\beta \Omega / 2}  \tilde{\mathfrak{W}}_{\bar{\nu} s'}(-\Omega, \vec{K}_\perp) \hat{D}_{\bar{\nu} s'}^\dagger(\Omega,-\vec{K}_\perp) \right] , \\
& \hat{d}_s^\dagger(\vec{k}) = \sum_{\nu=\{\text{L},\text{R}\}}  \sum_{s'=1}^2 \int_0^\infty d\Omega \int_{\mathbb{R}^2} d^2 K_\perp  \alpha^*_{\nu}(\vec{k},\Omega,\vec{K}_\perp)  \nonumber \\
& \times   \tilde{v}^\dagger_s(\vec{k}) \left[ -s_\nu i e^{-\beta \Omega / 2}  \tilde{\mathfrak{W}}_{\bar{\nu} s'}(\Omega, -\vec{K}_\perp)  \hat{C}_{\bar{\nu} s'}(\Omega,-\vec{K}_\perp)\right. \nonumber \\
&  \left. +   \tilde{\mathfrak{W}}_{\nu s'}(-\Omega, -\vec{K}_\perp) \hat{D}_{\nu s'}^\dagger(\Omega,\vec{K}_\perp) \right].
\end{align}
\end{subequations}
Finally, by using Eq.~(\ref{W_L_W_R}),
\begin{subequations}\label{Bogoliubov_transformations_3_Rindler_7}
\begin{align}
& \hat{c}_s(\vec{k}) = \sum_{\nu=\{\text{L},\text{R}\}}  \sum_{s'=1}^2 \int_0^\infty d\Omega \int_{\mathbb{R}^2} d^2 K_\perp   \alpha_{\nu}(\vec{k},\Omega,\vec{K}_\perp) \nonumber \\
& \times  \tilde{u}^\dagger_s(\vec{k}) \tilde{\mathfrak{W}}_{\nu s'}(\Omega, \vec{K}_\perp) \left[ \hat{C}_{\nu s'}(\Omega,\vec{K}_\perp) \right. \nonumber \\
&  \left. +  s_\nu i e^{-\beta \Omega / 2}  \sum_{s''=1}^2 M_{\nu s''s'}(\Omega, \vec{K}_\perp) \hat{D}_{\bar{\nu} s''}^\dagger(\Omega,-\vec{K}_\perp) \right] ,\label{Bogoliubov_transformations_3_Rindler_7_a} \\
& \hat{d}_s^\dagger(\vec{k}) = \sum_{\nu=\{\text{L},\text{R}\}}  \sum_{s'=1}^2 \int_0^\infty d\Omega \int_{\mathbb{R}^2} d^2 K_\perp  \alpha^*_{\nu}(\vec{k},\Omega,\vec{K}_\perp)  \nonumber \\
& \times   \tilde{v}^\dagger_s(\vec{k})\tilde{\mathfrak{W}}_{\nu s'}(-\Omega, -\vec{K}_\perp) \left[   \hat{D}_{\nu s'}^\dagger(\Omega,\vec{K}_\perp)  \right. \nonumber \\
&  \left. -s_\nu i e^{-\beta \Omega / 2}  \sum_{s''=1}^2 M_{\nu s''s'}(-\Omega, -\vec{K}_\perp)  \hat{C}_{\bar{\nu} s''}(\Omega,-\vec{K}_\perp)\right].\label{Bogoliubov_transformations_3_Rindler_7_b}
\end{align}
\end{subequations}

From the definition of $M_{\nu ss'}(\Omega, \vec{K}_\perp)$ [Eq.~(\ref{M})], one can compute its complex conjugate, that reads
\begin{equation}\label{M_conjugate}
M^*_{\nu ss'}(\Omega, \vec{K}_\perp) = M_{\bar{\nu} s's}(-\Omega, \vec{K}_\perp).
\end{equation}
By using Eq.~(\ref{M_conjugate}), one can also conjugate Eq.~(\ref{Bogoliubov_transformations_3_Rindler_7_b}) to obtain
\begin{align}\label{Bogoliubov_transformations_3_Rindler_7_b_conjugate}
& \hat{d}_s(\vec{k}) = \sum_{\nu=\{\text{L},\text{R}\}}  \sum_{s'=1}^2 \int_0^\infty d\Omega \int_{\mathbb{R}^2} d^2 K_\perp  \alpha_{\nu}(\vec{k},\Omega,\vec{K}_\perp)  \nonumber \\
& \times   \tilde{\mathfrak{W}}^\dagger_{\nu s'}(-\Omega, -\vec{K}_\perp) \tilde{v}_s(\vec{k})\left[   \hat{D}_{\nu s'}(\Omega,\vec{K}_\perp)  \right. \nonumber \\
&  \left. + s_\nu i e^{-\beta \Omega / 2}  \sum_{s''=1}^2 M_{\bar{\nu} s's''}(\Omega, -\vec{K}_\perp)  \hat{C}^\dagger_{\bar{\nu} s''}(\Omega,-\vec{K}_\perp)\right].
\end{align}

In conclusion, we computed the Bogoliubov transformations relating Minkowski and Rindler operators [Eq.~(\ref{Bogoliubov_transformations_3_Rindler_4})]. The explicit form of the Bogoliubov coefficient $\alpha_{\nu}(\vec{k},\Omega,\vec{K}_\perp)$ is reported in Eq.~(\ref{Bogoliubov_coefficient_2}). The symmetry between Bogoliubov coefficients of opposite wedge [Eq.~(\ref{alpha_beta_Unruh})] resulted in a coupling between Rindler operators of opposite wedge and frequency in the Bogoliubov transformation [Eqs.~(\ref{Bogoliubov_transformations_3_Rindler_7_a}) and (\ref{Bogoliubov_transformations_3_Rindler_7_b_conjugate})]. In the next section, we will show how this coupling is involved in the Rindler-Fock representation of the Minkowski vacuum.

\section{Minkowski vacuum in the left and right Rindler frame}\label{Minkowski_vacuum_in_Rindler_spacetime}

In Sec.~\ref{Bogoliubov_transformation} we derived the Bogoliubov transformations relating Minkowski and Rindler operators. We obtained an expression in which operators of opposite wedge and frequency are coupled. Here, we will use these transformations to show how the Minkowski vacuum can be represented as an element of the Rindler-Fock space. We will obtain two-modes squeezed states where each Rindler mode is paired with the mode with opposite wedge and frequency. The spin degrees of freedom are coupled through the matrix $M_{\nu ss'} (\Omega, \vec{K}_\perp)$ defined in Sec.~\ref{Rindler_Dirac_modes}. Hence, we will obtained different representations depending of the chosen basis $\tilde{\mathfrak{W}}_{\nu s}(\Omega, \vec{K}_\perp)$.

The Minkowski vacuum $|0_\text{M} \rangle $ is defined as the state that is always annihilated by the Minkowski operators $\hat{c}_s(\vec{k}) $ and $\hat{d}_s(\vec{k})$, i.e.,
\begin{align}\label{Minkowski_vacuum}
& \hat{c}_s(\vec{k}) |0_\text{M} \rangle = 0, & \hat{d}_s(\vec{k}) |0_\text{M} \rangle = 0,
\end{align}
for any $s$ and $\vec{k}$. Conversely, the Rindler vacuum $ |0_\text{L},0_\text{R} \rangle$ is defined in the following way
\begin{align}\label{Rindler_vacuum}
& \hat{C}_{\nu s}(\Omega,\vec{K}_\perp) |0_\text{L},0_\text{R} \rangle = 0, & \hat{D}_{\nu s}(\Omega,\vec{K}_\perp)|0_\text{L},0_\text{R} \rangle = 0,
\end{align}
for any $\nu$, $s$, $\Omega$ and $\vec{K}_\perp$.

In order to see $|0_\text{M} \rangle$ as an element of the Rindler-Fock space, one has to plug the Bogoliubov transformations (\ref{Bogoliubov_transformations_3_Rindler_7_a}) and (\ref{Bogoliubov_transformations_3_Rindler_7_b_conjugate}) in Eq.~(\ref{Minkowski_vacuum}) and look for a Rindler-Fock state such that
\begin{subequations}\label{Minkowski_vacuum_2}
\begin{align}
& \sum_{\nu=\{\text{L},\text{R}\}}  \sum_{s'=1}^2 \int_0^\infty d\Omega \int_{\mathbb{R}^2} d^2 K_\perp   \alpha_{\nu}(\vec{k},\Omega,\vec{K}_\perp) \nonumber \\
& \times  \tilde{u}^\dagger_s(\vec{k}) \tilde{\mathfrak{W}}_{\nu s'}(\Omega, \vec{K}_\perp) \left[ \hat{C}_{\nu s'}(\Omega,\vec{K}_\perp) +  s_\nu i e^{-\beta \Omega / 2} \right. \nonumber \\
&  \left. \times \sum_{s''=1}^2 M_{\nu s''s'}(\Omega, \vec{K}_\perp) \hat{D}_{\bar{\nu} s''}^\dagger(\Omega,-\vec{K}_\perp) \right]  |0_\text{M} \rangle = 0, \\
& \sum_{\nu=\{\text{L},\text{R}\}}  \sum_{s'=1}^2 \int_0^\infty d\Omega \int_{\mathbb{R}^2} d^2 K_\perp  \alpha_{\nu}(\vec{k},\Omega,\vec{K}_\perp)  \nonumber \\
& \times   \tilde{\mathfrak{W}}^\dagger_{\nu s'}(-\Omega, -\vec{K}_\perp) \tilde{v}_s(\vec{k})\left[   \hat{D}_{\nu s'}(\Omega,\vec{K}_\perp) + s_\nu i e^{-\beta \Omega / 2}  \right. \nonumber \\
&  \left. \times \sum_{s''=1}^2 M_{\bar{\nu} s's''}(\Omega, -\vec{K}_\perp)  \hat{C}^\dagger_{\bar{\nu} s''}(\Omega,-\vec{K}_\perp)\right] |0_\text{M} \rangle = 0, 
\end{align}
\end{subequations}
for any $s$ and $\vec{k}$. As a consequence of the generality of $s$ and $\vec{k}$, Eq.~(\ref{Minkowski_vacuum_2}) reads
\begin{subequations}\label{Minkowski_vacuum_Rindler}
\begin{align}
& \left[ \hat{C}_{\nu s}(\Omega,\vec{K}_\perp) +  s_\nu i e^{-\beta \Omega / 2} \sum_{s'=1}^2 M_{\nu s's}(\Omega, \vec{K}_\perp) \right. \nonumber \\
&  \left. \times  \hat{D}_{\bar{\nu} s'}^\dagger(\Omega,-\vec{K}_\perp) \right]  |0_\text{M} \rangle = 0,\label{Minkowski_vacuum_Rindler_a}\\
& \left[   \hat{D}_{\nu s}(\Omega,\vec{K}_\perp) + s_\nu i e^{-\beta \Omega / 2}  \sum_{s'=1}^2 M_{\bar{\nu} ss'}(\Omega, -\vec{K}_\perp)  \right. \nonumber \\
&  \left. \times \hat{C}^\dagger_{\bar{\nu} s'}(\Omega,-\vec{K}_\perp)\right] |0_\text{M} \rangle = 0.\label{Minkowski_vacuum_Rindler_b}
\end{align}
\end{subequations}

A solution for Eq.~(\ref{Minkowski_vacuum_Rindler}) exists and reads
\begin{equation}\label{0_M_0_LR}
|0_\text{M} \rangle \propto \exp (\hat{O} ) |0_\text{L},0_\text{R} \rangle,
\end{equation}
with
\begin{align}\label{O}
\hat{O} = & - i \sum_{\nu=\{\text{L},\text{R}\}} s_\nu \sum_{s=1}^2 \sum_{s'=1}^2 \int_0^{+\infty} d\Omega \int_{\mathbb{R}^2} d^2 K_\perp  e^{-\beta \Omega / 2}   \nonumber \\ 
&  \times  M_{\nu s's} (\Omega, \vec{K}_\perp) \hat{C}^\dagger_{\nu s}(\Omega,\vec{K}_\perp)\hat{D}^\dagger_{\bar{\nu} s'}(\Omega,-\vec{K}_\perp).
\end{align}
Equation (\ref{0_M_0_LR}) is the representation of the Minkowski vacuum in terms of left and right Rindler particles.

We now provide a proof for Eq.~(\ref{0_M_0_LR}) as the solution of Eq.~(\ref{Minkowski_vacuum_Rindler}). By using the anticommutation properties of the Dirac operators (\ref{anticommutating_rules}), one obtains the following identities
\begin{subequations}\label{anticommutating_rules_2}
\begin{align}
& \hat{C}_{\nu s}(\Omega,\vec{K}_\perp) \hat{C}^\dagger_{\nu' s'}(\Omega',\vec{K}'_\perp) \hat{D}^\dagger_{\bar{\nu}' s''}(\Omega',-\vec{K}'_\perp) \nonumber \\
= & \delta_{\nu \nu'} \delta_{ss'} \delta(\Omega-\Omega') \delta^2(\vec{K}_\perp-\vec{K}'_\perp)\hat{D}^\dagger_{\bar{\nu} s''}(\Omega,-\vec{K}_\perp)\nonumber \\
& + \hat{C}^\dagger_{\nu' s'}(\Omega',\vec{K}'_\perp) \hat{D}^\dagger_{\bar{\nu}' s''}(\Omega',-\vec{K}'_\perp) \hat{C}_{\nu s}(\Omega,\vec{K}_\perp), \\
& \hat{D}_{\nu s}(\Omega,\vec{K}_\perp) \hat{C}^\dagger_{\nu' s'}(\Omega',\vec{K}'_\perp) \hat{D}^\dagger_{\bar{\nu}' s''}(\Omega',-\vec{K}'_\perp) \nonumber \\
= & - \delta_{\nu \bar{\nu}'} \delta_{ss''} \delta(\Omega-\Omega') \delta^2(\vec{K}_\perp+\vec{K}'_\perp)\hat{C}^\dagger_{\bar{\nu} s'}(\Omega,-\vec{K}_\perp)\nonumber \\
& + \hat{C}^\dagger_{\nu' s'}(\Omega',\vec{K}'_\perp) \hat{D}^\dagger_{\bar{\nu}' s''}(\Omega',-\vec{K}'_\perp)\hat{D}_{\nu s}(\Omega,\vec{K}_\perp),  \\
& \hat{C}^\dagger_{\nu s}(\Omega,\vec{K}_\perp) \hat{C}^\dagger_{\nu' s'}(\Omega',\vec{K}'_\perp) \hat{D}^\dagger_{\bar{\nu}' s''}(\Omega',-\vec{K}'_\perp) \nonumber \\
= & \hat{C}^\dagger_{\nu' s'}(\Omega',\vec{K}'_\perp) \hat{D}^\dagger_{\bar{\nu}' s''}(\Omega',-\vec{K}'_\perp) \hat{C}^\dagger_{\nu s}(\Omega,\vec{K}_\perp),  \\
 &  \hat{D}^\dagger_{\nu s}(\Omega,\vec{K}_\perp) \hat{C}^\dagger_{\nu' s'}(\Omega',\vec{K}'_\perp) \hat{D}^\dagger_{\bar{\nu}' s''}(\Omega',-\vec{K}'_\perp)  \nonumber \\
= &  \hat{C}^\dagger_{\nu' s'}(\Omega',\vec{K}'_\perp) \hat{D}^\dagger_{\bar{\nu}' s''}(\Omega',-\vec{K}'_\perp) \hat{D}^\dagger_{\nu s}(\Omega,\vec{K}_\perp).
\end{align} 
\end{subequations}
By using Eqs.~(\ref{O}) and (\ref{anticommutating_rules_2}) and the fact that $s_{\bar{\nu}} = - s_\nu$,
\begin{subequations}\label{anticommutating_rules_3}
\begin{align}
&  \hat{C}_{\nu s}(\Omega,\vec{K}_\perp)\hat{O} = -s_\nu i  e^{-\beta \Omega / 2} \sum_{s'=1}^2 M_{\nu s's} (\Omega, \vec{K}_\perp)\nonumber \\
& \times \hat{D}^\dagger_{\bar{\nu} s'}(\Omega,-\vec{K}_\perp)+ \hat{O} \hat{C}_{\nu s}(\Omega,\vec{K}_\perp), \\
&  \hat{D}_{\nu s}(\Omega,\vec{K}_\perp)\hat{O}  = -s_\nu i  e^{-\beta \Omega / 2} \sum_{s'=1}^2 M_{\bar{\nu} ss'} (\Omega, -\vec{K}_\perp) \nonumber \\
& \times \hat{C}^\dagger_{\bar{\nu} s'}(\Omega,-\vec{K}_\perp) +\hat{O}   \hat{D}_{\nu s}(\Omega,\vec{K}_\perp) , \\
&  \hat{C}^\dagger_{\nu s}(\Omega,\vec{K}_\perp)\hat{O} =  \hat{O} \hat{C}^\dagger_{\nu s}(\Omega,\vec{K}_\perp), \\
&  \hat{D}^\dagger_{\nu s}(\Omega,\vec{K}_\perp)\hat{O}  = \hat{O}   \hat{D}^\dagger_{\nu s}(\Omega,\vec{K}_\perp) .
\end{align}
\end{subequations}
Recursively one may prove the following identity from Eq.~(\ref{anticommutating_rules_3})
\begin{subequations}\label{anticommutating_rules_4}
\begin{align}
& \hat{C}_{\nu s}(\Omega,\vec{K}_\perp) \hat{O}^n =  -n s_\nu i e^{-\beta \Omega / 2} \sum_{s'=1}^2 M_{\nu s's} (\Omega, \vec{K}_\perp) \nonumber \\
& \times \hat{D}^\dagger_{\bar{\nu} s'}(\Omega,-\vec{K}_\perp) \hat{O}^{n-1} + \hat{O}^n \hat{C}_{\nu s}(\Omega,\vec{K}_\perp) , \\
& \hat{D}_{\nu s}(\Omega,\vec{K}_\perp) \hat{O}^n = -n s_\nu i  e^{-\beta \Omega / 2}  \sum_{s'=1}^2 M_{\bar{\nu} ss'} (\Omega, -\vec{K}_\perp) \nonumber \\
& \times \hat{C}^\dagger_{\bar{\nu} s'}(\Omega,-\vec{K}_\perp) \hat{O}^{n-1} + \hat{O}^n \hat{D}_{\nu s}(\Omega,\vec{K}_\perp),
\end{align}
\end{subequations}
which holds for any $n \in \mathbb{N}$. By acting on the Rindler vacuum state $|0_\text{L},0_\text{R} \rangle$, Eq.~(\ref{anticommutating_rules_4}) leads to
\begin{subequations}\label{anticommutating_rules_5}
\begin{align}
& \hat{C}_{\nu s}(\Omega,\vec{K}_\perp) \hat{O}^n |0_\text{L},0_\text{R} \rangle = -n s_\nu i e^{-\beta \Omega / 2} \nonumber \\
 & \times \sum_{s'=1}^2 M_{\nu s's} (\Omega, \vec{K}_\perp)  \hat{D}^\dagger_{\bar{\nu} s'}(\Omega,-\vec{K}_\perp) \hat{O}^{n-1}|0_\text{L},0_\text{R} \rangle , \\
& \hat{D}_{\nu s}(\Omega,\vec{K}_\perp) \hat{O}^n|0_\text{L},0_\text{R} \rangle = -n s_\nu i  e^{-\beta \Omega / 2} \nonumber \\
 & \times \sum_{s'=1}^2 M_{\bar{\nu} ss'} (\Omega, -\vec{K}_\perp) \hat{C}^\dagger_{\bar{\nu} s'}(\Omega,-\vec{K}_\perp) \hat{O}^{n-1}|0_\text{L},0_\text{R} \rangle.
\end{align}
\end{subequations}
By multiplying Eq.~(\ref{anticommutating_rules_5}) with $1/n!$ and summing with respect to $n$, one obtains
\begin{subequations}\label{anticommutating_rules_6}
\begin{align}
& \hat{C}_{\nu s}(\Omega,\vec{K}_\perp) \exp (\hat{O} ) |0_\text{L},0_\text{R} \rangle =  -s_\nu i e^{-\beta \Omega / 2}  \nonumber \\
& \times \sum_{s'=1}^2 M_{\nu s's} (\Omega, \vec{K}_\perp) \hat{D}^\dagger_{\bar{\nu} s'}(\Omega,-\vec{K}_\perp) \exp (\hat{O} )|0_\text{L},0_\text{R} \rangle ,\\
& \hat{D}_{\nu s}(\Omega,\vec{K}_\perp) \exp (\hat{O} )|0_\text{L},0_\text{R} \rangle  =  -s_\nu i  e^{-\beta \Omega / 2}\nonumber \\
& \times  \sum_{s'=1}^2 M_{\bar{\nu} ss'} (\Omega, -\vec{K}_\perp) \hat{C}^\dagger_{\bar{\nu} s'}(\Omega,-\vec{K}_\perp)  \exp (\hat{O} )|0_\text{L},0_\text{R} \rangle,
\end{align}
\end{subequations}
which proves that Eq.~(\ref{0_M_0_LR}) is the solution of Eq.~(\ref{Minkowski_vacuum_Rindler}).

We now show how to write Eq.~(\ref{O}) in a more compact form. By computing the sum with respect to $\nu$ and performing the integral variables transformation $\Omega \mapsto -\Omega $ and $\vec{K}_\perp \mapsto -\vec{K}_\perp $ when $\nu = \text{L}$, one obtains
\begin{align}\label{O_2}
\hat{O} = & - i \sum_{s=1}^2 \sum_{s'=1}^2 \left[ - \int_{-\infty}^0 d\Omega \int_{\mathbb{R}^2} d^2 K_\perp  e^{\beta \Omega / 2} \right.  \nonumber \\ 
&  \times  M_{\text{L} s's} (-\Omega, -\vec{K}_\perp) \hat{C}^\dagger_{\text{L} s}(-\Omega,-\vec{K}_\perp)\hat{D}^\dagger_{\text{R} s'}(-\Omega,\vec{K}_\perp)  \nonumber \\ 
&  + \int_0^{+\infty} d\Omega \int_{\mathbb{R}^2} d^2 K_\perp  e^{-\beta \Omega / 2} M_{\text{R}  s's} (\Omega, \vec{K}_\perp)   \nonumber \\ 
& \left. \times  \hat{C}^\dagger_{\text{R} s}(\Omega,\vec{K}_\perp)\hat{D}^\dagger_{\text{L} s'}(\Omega,-\vec{K}_\perp) \right].
\end{align}
By letting $\hat{C}^\dagger_{\text{R} s}(\Omega,\vec{K}_\perp)$ and $\hat{D}^\dagger_{\text{L} s'}(\Omega,-\vec{K}_\perp)$ anticommute [Eq.~(\ref{anticommutating_rules_e})],
\begin{align}\label{O_3}
\hat{O} = & i \sum_{s=1}^2 \sum_{s'=1}^2 \left[ \int_{-\infty}^0 d\Omega \int_{\mathbb{R}^2} d^2 K_\perp  e^{\beta \Omega / 2} \right.  \nonumber \\ 
&  \times  M_{\text{L} s's} (-\Omega, -\vec{K}_\perp) \hat{C}^\dagger_{\text{L} s}(-\Omega,-\vec{K}_\perp)\hat{D}^\dagger_{\text{R} s'}(-\Omega,\vec{K}_\perp)  \nonumber \\ 
&  + \int_0^{+\infty} d\Omega \int_{\mathbb{R}^2} d^2 K_\perp  e^{-\beta \Omega / 2}  M_{\text{R}  s's} (\Omega, \vec{K}_\perp)   \nonumber \\ 
& \left. \times \hat{D}^\dagger_{\text{L} s'}(\Omega,-\vec{K}_\perp) \hat{C}^\dagger_{\text{R} s}(\Omega,\vec{K}_\perp) \right].
\end{align}

Equation (\ref{O_3}) suggests the definition of the operators $\hat{E}(\Theta)$, with $\Theta = (\nu, s, \Omega,\vec{K}_\perp) \in  \{ \text{L}, \text{R} \} \otimes \{ 1, 2\} \otimes \mathbb{R}^3$, such that
\begin{subequations}\label{E_CD}
\begin{align}
& \hat{E}^\dagger(\text{L}, s, \Omega,\vec{K}_\perp) = \nonumber \\  &  \begin{cases}
\sum_{s'=1}^2  M_{\text{R} s's} (\Omega,\vec{K}_\perp) \hat{D}^\dagger_{\text{L} s'}(\Omega,-\vec{K}_\perp) & \text{if } \Omega>0 \\
\sum_{s'=1}^2 M_{\text{L} ss'} (-\Omega,-\vec{K}_\perp) \hat{C}^\dagger_{\text{L} s'}(-\Omega,-\vec{K}_\perp)  & \text{if } \Omega<0
\end{cases},\label{E_CD_a} \\
& \hat{E}^\dagger(\text{R}, s, \Omega,\vec{K}_\perp) = \begin{cases}
\hat{C}^\dagger_{\text{R} s}(\Omega,\vec{K}_\perp)  & \text{if } \Omega>0\\
\hat{D}^\dagger_{\text{R} s}(-\Omega,\vec{K}_\perp) & \text{if } \Omega<0
\end{cases}.\label{E_CD_b}
\end{align}
\end{subequations}
In this way, Eq.~(\ref{O_3}) reads
\begin{equation}\label{O_4}
\hat{O} = \sum_\theta  f(\theta) \hat{F}^\dagger(\theta),
\end{equation}
with $\theta = (s, \Omega,\vec{K}_\perp) \in  \{ 1, 2\} \otimes \mathbb{R}^3$, 
\begin{subequations}
\begin{align}
& f(s, \Omega,\vec{K}_\perp) = i e^{-\beta |\Omega| / 2} ,\label{f} \\
 & \hat{F}^\dagger(\theta) = \hat{E}^\dagger(\text{L}, \theta) \hat{E}^\dagger(\text{R}, \theta) \label{F}
\end{align}
\end{subequations}
and where $\sum_\theta$ is a generalized sum for the $\theta$ variables consisting in a sum with respect to the discrete variable $s$ and an integral for the continuum variables $\Omega$ and $\vec{K}_\perp$, i.e.,
\begin{align}
& \sum_\theta = \sum_{s=1}^2 \int_{\mathbb{R}} d \Omega \int_{\mathbb{R}^2} d^2 K_\perp, & \theta = (s, \Omega,\vec{K}_\perp).
\end{align}

Since the matrix $M_{\nu ss'}(\Omega, \vec{K}_\perp)$ is unitary, Eq.~(\ref{E_CD}) is invertible. Indeed, by using  Eqs.~(\ref{M_unitary}) and (\ref{E_CD}) one can prove the following identities
\begin{subequations}\label{CD_E}
\begin{align}
& \hat{C}^\dagger_{\text{L} s}(\Omega,\vec{K}_\perp) = \sum_{s'=1}^2 M^*_{\text{L} s's} (\Omega,\vec{K}_\perp) \hat{E}^\dagger(\text{L}, s',-\Omega,-\vec{K}_\perp), \\
 & \hat{D}^\dagger_{\text{L} s}(\Omega,\vec{K}_\perp) = \sum_{s'=1}^2 M^*_{\text{R} ss'} (\Omega,-\vec{K}_\perp) \hat{E}^\dagger(\text{L}, s', \Omega,-\vec{K}_\perp),\\
& \hat{C}^\dagger_{\text{R} s}(\Omega,\vec{K}_\perp) = \hat{E}^\dagger(\text{R}, s , \Omega,\vec{K}_\perp),\label{CD_E_c} \\
 & \hat{D}^\dagger_{\text{R} s}(\Omega,\vec{K}_\perp) = \hat{E}^\dagger(\text{R}, s,- \Omega,\vec{K}_\perp),\label{CD_E_d} 
\end{align}
\end{subequations}
for any $\Omega > 0$. For this reason, Eqs.~(\ref{E_CD}) and (\ref{CD_E}) are a one-to-one mapping between $\hat{E}^\dagger(\Theta)$ and the creation operators $\hat{C}^\dagger_{\nu s}(\Omega,\vec{K}_\perp)$ and $\hat{D}^\dagger_{\nu s}(\Omega,\vec{K}_\perp)$.

Notice that, from the definition of the Rindler vacuum $|0_\text{L},0_\text{R} \rangle$ [Eq.~(\ref{Rindler_vacuum})] and the operator $\hat{E}(\Theta)$ [Eq.~(\ref{E_CD})],
\begin{equation} \label{E_annhilator}
\hat{E}(\Theta)|0_\text{L},0_\text{R} \rangle = 0.
\end{equation}
The anticommutation properties for the operator $\hat{E}(\Theta)$ read
\begin{subequations}\label{anticommutating_rules_E}
\begin{align}
& \{ \hat{E}(\Theta), \hat{E}^\dagger(\Theta') \} =  \Delta(\Theta,\Theta'),  \label{anticommutating_rules_E_a}\\
&  \{ \hat{E}(\Theta), \hat{E}(\Theta') \}  = 0 \label{anticommutating_rules_E_b},
\end{align}
\end{subequations}
where $\Delta(\Theta,\Theta')$ is a generalized delta function for the variables $\Theta = (\nu, s, \Omega,\vec{K}_\perp)$ and $\Theta' = (\nu', s', \Omega',\vec{K}'_\perp)$. The function $\Delta(\Theta,\Theta')$ is the product of the Kronecker delta for the discrete variables $\nu$, $\nu'$, $s$ and $s'$ and the Dirac delta function for the continuum variables $\Omega-\Omega'$ and $\vec{K}_\perp - \vec{K}'_\perp$:
\begin{align}
& \Delta((\nu, s, \Omega,\vec{K}_\perp), (\nu', s', \Omega',\vec{K}'_\perp)) \nonumber \\
 = & \delta_{\nu\nu'} \delta_{ss'} \delta(\Omega-\Omega') \delta^2(\vec{K}_\perp - \vec{K}'_\perp).
\end{align}
Equation (\ref{anticommutating_rules_E}) can be checked by using Eqs.~(\ref{anticommutating_rules}), (\ref{M_unitary}) and (\ref{E_CD}). As a consequence of Eqs.~(\ref{E_annhilator}) and (\ref{anticommutating_rules_E}), the mapping from $\hat{C}^\dagger_{\nu s}(\Omega,\vec{K}_\perp)$ and $\hat{D}^\dagger_{\nu s}(\Omega,\vec{K}_\perp)$ to $\hat{E}^\dagger(\Theta)$ is canonical.

Equations (\ref{F}) and (\ref{anticommutating_rules_E_b}) lead to
\begin{subequations}\label{commutating_rules_F}
\begin{align}
& [ \hat{F}^\dagger(\theta), \hat{F}^\dagger(\theta') ] = 0,  \label{commutating_rules_F_a}\\
&  \hat{F}^\dagger(\theta) \hat{F}^\dagger(\theta) = 0 \label{commutating_rules_F_b}.
\end{align}
\end{subequations}
The representative of the Minkowski vacuum given by Eqs.~(\ref{0_M_0_LR}) and (\ref{O_4}) and the algebraic properties of the operators $\hat{E}(\Theta)$ and $\hat{F}(\theta)$ [Eqs.~(\ref{E_annhilator}), (\ref{anticommutating_rules_E}) and (\ref{commutating_rules_F})] will be used in the next section to derive the statistical operator describing the Minkowski vacuum in the right Rindler frame.

\section{Minkowski vacuum in the right Rindler frame}\label{Unruh_effect_for_Dirac_fields}

In the previous section we derived the representation of the Minkowski vacuum in both left and right Rindler frames. Here, instead, we will focus only on the right wedge, which describes the accelerated observer with positive acceleration $c^2 a$. By performing a partial trace over the left wedge, we will compute the statistical operator representing the Minkowski vacuum as an element of the right Rindler-Fock space. The result will be a fermionic thermal state, which is at the origin of the Unruh effect for Dirac fields.

In order to perform the partial trace, one needs a basis for the Rindler-Fock space. The single particle space is defined by the creation operators $\hat{C}^\dagger_{\nu s}(\Omega,\vec{K}_\perp)$ and $\hat{D}^\dagger_{\nu s}(\Omega,\vec{K}_\perp)$ acting on the vacuum state $|0_\text{L},0_\text{R} \rangle$. Hence, a basis for single particles and antiparticles in each wedge can be defined through the quantum numbers $s$, $\Omega$ and $\vec{K}_\perp$. Alternatively, one may take advantage of the canonical transformation (\ref{E_CD}) and use the operator $\hat{E}(\Theta)$ and the quantum numbers $\Theta = (\nu, s, \Omega,\vec{K}_\perp)$ to describe single particles and antiparticles of both wedges in the following way
\begin{equation}\label{single_particle_basis}
| \Theta \rangle  = \hat{E}^\dagger(\Theta) |0_\text{L},0_\text{R} \rangle.
\end{equation}
Notice that Eq.~(\ref{single_particle_basis}) is an orthonormal basis for the single particle and antiparticle space. The orthonormality condition can be checked by using Eqs.~(\ref{E_annhilator}) and (\ref{anticommutating_rules_E_a}).

Many-particles states are given by the action of sequences of creation operators $\hat{E}^\dagger(\Theta)$ on the Rindler vacuum. We define the following Rindler-Fock state
\begin{equation}\label{many_particle_Rindler_state}
| \mathbf{\Theta} \rangle = \hat{\mathbf{E}}^\dagger (\mathbf{\Theta}) |0_\text{L},0_\text{R} \rangle,
\end{equation}
with
\begin{equation}\label{EE}
\hat{\mathbf{E}}^\dagger (\mathbf{\Theta}) = \prod_{i=1}^{|\mathbf{\Theta}|} \hat{E}^\dagger(\Theta_i)
\end{equation}
and where $\mathbf{\Theta} = \{ \Theta_1, \dots, \Theta_n \}$ is an ordered set of quantum numbers $\Theta_i$ and $|\mathbf{\Theta}|$ the cardinality of the set. By using Eqs.~(\ref{E_annhilator}) and (\ref{anticommutating_rules_E}), one can prove that the scalar product of different states defined by Eq.~(\ref{many_particle_Rindler_state}) reads
\begin{equation}\label{many_particle_Rindler_state_product}
 \langle  \mathbf{\Theta} | \mathbf{\Theta}' \rangle =  \sum_{\tau \in S_{|\mathbf{\Theta}|}} \text{sign}(\tau)  \mathbf{\Delta}(\tau(\mathbf{\Theta}), \mathbf{\Theta}'),
\end{equation}
with
\begin{equation}\label{DDelta}
\mathbf{\Delta}(\mathbf{\Theta}, \mathbf{\Theta}') = \delta_{|\mathbf{\Theta}||\mathbf{\Theta}'|}  \prod_{i=1}^{|\mathbf{\Theta}|} \Delta\left( \Theta_i,\Theta'_{i} \right) 
\end{equation}
and where $S_n$ is the space of all permutations of sets with $n$ elements.

Notice that the order of the creation operators $\hat{E}^\dagger(\Theta_i)$ on the right side of Eq.~(\ref{many_particle_Rindler_state}) cannot be ignored because of the anticommuting nature of the Rindler operators $\hat{E}^\dagger(\Theta_i)$ [Eq.~(\ref{anticommutating_rules_E_b})]. Any permutation of quantum numbers $\Theta_i$ leads to the same many-particles state up to a sign. The set of states $| \mathbf{\Theta} \rangle$ cannot be chosen as basis, due to the presence of the sign of permutations appearing in Eq.~(\ref{many_particle_Rindler_state_product}).

To define a basis for the particles space, one has to consider an operator $\mathcal{O}$ that acts on any sequence of quantum numbers $\mathbf{\Theta}$ and rearrange their order by following a fixed ordering rule. The set of states $| \mathcal{O} (\mathbf{\Theta}) \rangle$ form an orthonormal basis for the many-particles space. Indeed the following equation holds
\begin{equation}\label{many_particle_Rindler_basis_product}
 \langle  \mathcal{O} (\mathbf{\Theta}) | \mathcal{O} (\mathbf{\Theta}') \rangle =  \sum_{\tau \in S_{|\mathbf{\Theta}|}}  \mathbf{\Delta}(\tau(\mathbf{\Theta}), \mathbf{\Theta}').
\end{equation}
Notice that in Eq.~(\ref{many_particle_Rindler_basis_product}) the sign of permutations is absent, as opposed to Eq.~(\ref{many_particle_Rindler_state_product}).

The orthonormality condition (\ref{many_particle_Rindler_basis_product}) can be proven in the following way. Firstly notice that the ordering function $\mathcal{O}$ acts on any sequence of quantum numbers $\mathbf{\Theta}$ as a $\mathbf{\Theta}$ dependent permutation. Indeed, for any $\mathbf{\Theta}$, one can define a permutation $\mathcal{P}_\mathbf{\Theta} \in S_{|\mathbf{\Theta}|}$ such that
\begin{equation}\label{P_Theta}
\mathcal{O} (\mathbf{\Theta}) = \mathcal{P}_\mathbf{\Theta} (\mathbf{\Theta}).
\end{equation}
Notice also that the ordering function $\mathcal{O}$ is unaffected by any permutation. Explicitly, this means that
\begin{equation} \label{O_ordering_permutation}
 \mathcal{O} ( \tau(\mathbf{\Theta}) ) = \mathcal{O} (\mathbf{\Theta}),
\end{equation}
for any $\tau \in S_{|\mathbf{\Theta}|}$. By using Eq.~(\ref{P_Theta}) in Eq.~(\ref{O_ordering_permutation}), one can also write
\begin{equation} \label{O_ordering_permutation_2}
\mathcal{P}_\mathbf{\tau(\Theta)} \tau(\mathbf{\Theta}) = \mathcal{P}_\mathbf{\Theta} (\mathbf{\Theta}),
\end{equation}
which means that
\begin{equation}\label{O_ordering_permutation_3}
\text{sign} ( \mathcal{P}_\mathbf{\tau(\Theta)} \tau ) = \text{sign} (\mathcal{P}_\mathbf{\Theta}).
\end{equation}

Equations (\ref{many_particle_Rindler_state_product}) and (\ref{P_Theta}) lead to the following scalar product
\begin{equation}\label{many_particle_Rindler_state_product_2}
 \langle  \mathcal{O}(\mathbf{\Theta}) | \mathcal{O}(\mathbf{\Theta}') \rangle  = \sum_{\tau \in S_{|\mathbf{\Theta}|}} \text{sign}(\tau)   \mathbf{\Delta}(\tau \mathcal{P}_\mathbf{\Theta}(\mathbf{\Theta}), \mathcal{P}_{\mathbf{\Theta}'}(\mathbf{\Theta}')).
\end{equation}
Notice that from the definition of $\mathbf{\Delta}(\mathbf{\Theta}, \mathbf{\Theta}')$ [Eq.~(\ref{DDelta})], by rearranging the order of the product index $i \mapsto \tau(i)$ with any permutation $\tau \in S_{|\mathbf{\Theta}|}$ one obtains
\begin{equation}
\mathbf{\Delta}(\tau(\mathbf{\Theta}), \tau(\mathbf{\Theta}')) = \mathbf{\Delta}(\mathbf{\Theta}, \mathbf{\Theta}').
\end{equation}
This can be used in Eq.~(\ref{many_particle_Rindler_state_product_2}) to obtain
\begin{equation}\label{many_particle_Rindler_state_product_3}
\langle  \mathcal{O}(\mathbf{\Theta}) | \mathcal{O}(\mathbf{\Theta}') \rangle  =  \sum_{\tau \in S_{|\mathbf{\Theta}|}} \text{sign}(\tau)   \mathbf{\Delta}(\tau\mathcal{P}_{\mathbf{\Theta}}\mathcal{P}^{-1}_{\mathbf{\Theta}'}(\mathbf{\Theta}), \mathbf{\Theta}').
\end{equation}
By using the fact that the sum $\sum_{\tau \in S_{|\mathbf{\Theta}|}}$ runs over all permutations of $S_{|\mathbf{\Theta}|}$, one can perform the transformation $\tau \mapsto \tau \mathcal{P}_{\mathbf{\Theta}'} \mathcal{P}^{-1}_{\mathbf{\Theta}}$ in Eq.~(\ref{many_particle_Rindler_state_product_3}) and write
\begin{equation}\label{many_particle_Rindler_state_product_4}
\langle  \mathcal{O}(\mathbf{\Theta}) | \mathcal{O}(\mathbf{\Theta}') \rangle = \sum_{\tau \in S_{|\mathbf{\Theta}|}} \text{sign}(\tau \mathcal{P}_{\mathbf{\Theta}'} \mathcal{P}^{-1}_{\mathbf{\Theta}})   \mathbf{\Delta}(\tau(\mathbf{\Theta}), \mathbf{\Theta}').
\end{equation}
Notice that the $\mathbf{\Delta}(\tau(\mathbf{\Theta}), \mathbf{\Theta}')$ function in the right side of Eq.~(\ref{many_particle_Rindler_state_product_4}) is nonvanishing only when $\mathbf{\Theta}' =  \tau(\mathbf{\Theta})$. Hence, Eq.~(\ref{many_particle_Rindler_state_product_4}) reads
\begin{equation}\label{many_particle_Rindler_state_product_5}
 \langle  \mathcal{O}(\mathbf{\Theta}) | \mathcal{O}(\mathbf{\Theta}') \rangle = \sum_{\tau \in S_{|\mathbf{\Theta}|}} \text{sign}(\tau \mathcal{P}_{\tau(\mathbf{\Theta})} \mathcal{P}^{-1}_{\mathbf{\Theta}})   \mathbf{\Delta}(\tau(\mathbf{\Theta}), \mathbf{\Theta}').
\end{equation}
By using Eq.~(\ref{O_ordering_permutation_3}) in (\ref{many_particle_Rindler_state_product_5}), one obtains Eq.~(\ref{many_particle_Rindler_basis_product}).

Equation (\ref{many_particle_Rindler_basis_product}) is the orthonormality condition for the many-particles states $| \mathcal{O} (\mathbf{\Theta}) \rangle$ defined as follows
\begin{equation}\label{many_particle_Rindler_basis}
| \mathcal{O} (\mathbf{\Theta}) \rangle =  \hat{\mathbf{E}}^\dagger (\mathcal{O}(\mathbf{\Theta})) |0_\text{L},0_\text{R} \rangle.
\end{equation}
Notice that Eq.~(\ref{many_particle_Rindler_basis}) is symmetric with respect to any permutation of the quantum numbers $\vec{\Theta}_i$ [Eq.~(\ref{O_ordering_permutation})].

Hereafter, we choose any ordering function $\mathcal{O}$ such that for any couple of quantum numbers $\Theta = (\nu, \theta)$ and $\Theta' = (\nu', \theta')$,
\begin{equation}\label{sorting_function}
\mathcal{O} ( \{ (\nu, \theta) , (\nu', \theta') \}  )= \begin{cases}
\mathcal{Q} ( \{ (\nu, \theta) , (\nu', \theta') \}  ) & \text{if } \theta \neq \theta' \\
\mathcal{W} ( \{ (\nu, \theta) , (\nu', \theta') \}  ) & \text{if } \theta = \theta'
 \end{cases},
\end{equation}
where $\mathcal{Q}$ is any ordering function with respect to the non-repeating quantum numbers $\theta = (s, \Omega, \vec{K}_\perp)$. The ordering function $\mathcal{W}$, instead, is with respect to the wedge variable $\nu$. We choose the following definition for $\mathcal{W}$
\begin{subequations} \label{W_sorting}
\begin{align}
& \mathcal{W} ( \{ (\text{L}, \theta) , (\text{R}, \theta) \}  ) =  ( \{ (\text{L}, \theta) , (\text{R}, \theta) \}  ), \\
& \mathcal{W} ( \{ (\text{R}, \theta) , (\text{L}, \theta) \}  ) =  ( \{ (\text{L}, \theta) , (\text{R}, \theta) \}  ).
\end{align}
\end{subequations}
We do not choose any particular definition for $\mathcal{Q}$. However, for completeness, we give a possible example by considering the lexicographical order as follows
\begin{subequations}
\begin{align}
& \mathcal{Q} ( \{ (\nu, s, \Omega, \vec{K}_\perp) , (\nu', s', \Omega', \vec{K}'_\perp) \}  ) \nonumber \\
 = & \begin{cases} 
 \{ (\nu, s, \Omega, \vec{K}_\perp) , (\nu', s', \Omega', \vec{K}'_\perp) \} & \text{if } s < s' \\
 \{ (\nu', s', \Omega', \vec{K}'_\perp) , (\nu, s, \Omega, \vec{K}_\perp) \} & \text{if } s > s'
 \end{cases}, \\
& \mathcal{Q} ( \{ (\nu, s, \Omega, \vec{K}_\perp) , (\nu', s, \Omega', \vec{K}'_\perp) \}  ) \nonumber \\
 = & \begin{cases} 
 \{ (\nu, s, \Omega, \vec{K}_\perp) , (\nu', s, \Omega', \vec{K}'_\perp) \} & \text{if } \Omega < \Omega' \\
 \{ (\nu', s, \Omega', \vec{K}'_\perp) , (\nu, s, \Omega, \vec{K}_\perp) \} & \text{if } \Omega > \Omega'
 \end{cases}, \\
& \mathcal{Q} ( \{ (\nu, s, \Omega, K_1, K_2) , (\nu', s, \Omega, K'_1, K'_2) \}  ) \nonumber \\
 = & \begin{cases} 
 \{ (\nu, s, \Omega, K_1, K_2) , (\nu', s, \Omega, K'_1, K'_2) \} & \text{if } K_1 < K'_1 \\
 \{ (\nu', s, \Omega, K'_1, K'_2) , (\nu, s, \Omega, K_1, K_2) \} & \text{if } K_1 > K'_1
 \end{cases}, \\
& \mathcal{Q} ( \{ (\nu, s, \Omega, K_1, K_2) , (\nu', s, \Omega, K_1, K'_2) \}  ) \nonumber \\
 = & \begin{cases} 
 \{ (\nu, s, \Omega, K_1, K_2) , (\nu', s, \Omega, K_1, K'_2) \} & \text{if } K_2 < K'_2 \\
 \{ (\nu', s, \Omega, K_1, K'_2) , (\nu, s, \Omega, K_1, K_2) \} & \text{if } K_2 > K'_2
 \end{cases}.
\end{align}
\end{subequations}

We now show how to write the Minkowski vacuum [Eq.~(\ref{0_M_0_LR})] in terms of the many-particles basis (\ref{many_particle_Rindler_basis}). Equation (\ref{O_4}) leads to
\begin{equation}\label{O_n}
\hat{O}^n = \sum_{\theta_1} \dots \sum_{\theta_n}  \prod_{i=1}^n f(\theta_i) \prod_{i=1}^n \hat{F}^\dagger(\theta_i),
\end{equation}
for any $n \in \mathbb{N}$. The operators $\hat{F}^\dagger(\theta_i)$ that appear in Eq.~(\ref{O_n}) are defined by Eq.~(\ref{F}) and can be written in terms of $\hat{\mathbf{E}}^\dagger (\mathbf{\Theta})$ [Eq.~(\ref{EE})] as follows
\begin{equation}\label{F_EE}
\hat{F}^\dagger(\theta) = \hat{\mathbf{E}}^\dagger ( \{ (\text{L}, \theta), (\text{R}, \theta) \} ).
\end{equation}
Notice that the couple of quantum numbers appearing in Eq.~(\ref{F_EE}) follow the $\mathcal{W}$ order [Eq.~(\ref{W_sorting})]. This means that
\begin{equation}\label{F_EE_2}
\hat{F}^\dagger(\theta) = \hat{\mathbf{E}}^\dagger ( \mathcal{W} ( \{ (\text{L}, \theta), (\text{R}, \theta) \} ) ).
\end{equation}

Consider the chain of operators $\prod_{i=1}^n \hat{F}^\dagger(\theta_i)$ that appears in Eq.~(\ref{O_n}). The operators $\hat{F}^\dagger(\theta_i)$ commute [Eq.~(\ref{commutating_rules_F_a})], and, hence, one may write Eq.~(\ref{O_n}) by following any order for the sequence of $\hat{F}^\dagger(\theta_i)$. Notice also that as a consequence of Eq.~(\ref{commutating_rules_F_b}), no repetition of the quantum numbers $\theta_i$ occurs. Therefore one may choose the $\mathcal{Q}$ order for the sequence of $\hat{F}^\dagger(\theta_i)$. By sorting the $\hat{F}^\dagger(\vec{\Theta}_i)$ operators in Eq.~(\ref{O_n}) with respect to the $\mathcal{Q}$ order and by considering the fact that the $\hat{E}^\dagger(\Theta)$ operators appearing in Eq.~(\ref{F}) already follow the $\mathcal{W}$ order [Eq.~(\ref{F_EE_2})], one derives the following identity
\begin{equation}\label{F_EE_3}
\prod_{i=1}^n \hat{F}^\dagger(\theta_i) = \hat{\mathbf{E}}^\dagger \left( \mathcal{O} \left( \bigcup_{i=1}^n \{ (\text{L}, \theta_i), (\text{R}, \theta_i) \} \right) \right).
\end{equation}

By plugging Eq.~(\ref{F_EE_3}) in Eq.~(\ref{O_n}), one obtains
\begin{align}\label{O_n_2}
\hat{O}^n = & \sum_{\theta_1} \dots \sum_{\theta_n}  \prod_{i=1}^n f(\theta_i) \nonumber \\
& \times \hat{\mathbf{E}}^\dagger \left( \mathcal{O} \left( \bigcup_{i=1}^n \{ (\text{L}, \theta_i), (\text{R}, \theta_i) \} \right) \right).
\end{align}
By acting on the Rindler vacuum and by using Eq.~(\ref{many_particle_Rindler_basis}), Eq.~(\ref{O_n_2}) reads
\begin{align}\label{O_n_3}
\hat{O}^n |0_\text{L},0_\text{R} \rangle = & \sum_{\theta_1} \dots \sum_{\theta_n}  \prod_{i=1}^n f(\theta_i) \nonumber \\
 & \times  \left| \mathcal{O} \left( \bigcup_{i=1}^n \{ (\text{L}, \theta_i), (\text{R}, \theta_i) \} \right)\right\rangle .
\end{align}
By multiplying Eq.~(\ref{O_n_3}) with $1/n!$ and summing with respect to $n$, one obtains
\begin{align}\label{exp_O_2}
\exp(\hat{O}) |0_\text{L},0_\text{R} \rangle =  &  |0_\text{L},0_\text{R} \rangle + \sum_{n=1}^\infty  \frac{1}{n!} \sum_{\theta_1} \dots \sum_{\theta_n}  \prod_{i=1}^n f(\theta_i) \nonumber \\
 & \times  \left| \mathcal{O} \left( \bigcup_{i=1}^n \{ (\text{L}, \theta_i), (\text{R}, \theta_i) \} \right)\right\rangle,
\end{align}
which provides a representation for the Minkowski vacuum [Eq.~(\ref{0_M_0_LR})] with respect to the basis (\ref{many_particle_Rindler_basis}).

We now compute the partial trace with respect to the left wedge. From Eq.~(\ref{exp_O_2}), one obtains
\begin{align}\label{tr_L_exp_O}
& \text{Tr}_\text{L} \left[ \exp(\hat{O}) | 0_\text{L},0_\text{R} \rangle \langle 0_\text{L},0_\text{R} | \exp(\hat{O})^\dagger \right]  = |0_\text{R} \rangle \langle 0_\text{R} | \nonumber \\
 & + \sum_{n=1}^\infty \sum_{m=1}^\infty \frac{1}{n!m!} \sum_{\theta_1} \dots \sum_{\theta_n} \sum_{\theta'_1} \dots \sum_{\theta'_m}  \prod_{i=1}^n f(\theta_i)  \prod_{i=1}^m f^*(\theta'_i)   \nonumber \\
 & \times \left\langle  \mathcal{O} \left( \bigcup_{i=1}^n \{ (\text{L}, \theta_i) \} \right) \left|  \mathcal{O} \left( \bigcup_{i=1}^m \{ (\text{L}, \theta'_i) \} \right) \right. \right\rangle  \nonumber \\
& \times  \left| \mathcal{O} \left( \bigcup_{i=1}^n \{ (\text{R}, \theta_i) \} \right) \right\rangle \left\langle \mathcal{O} \left( \bigcup_{i=1}^m \{ (\text{R}, \theta'_i) \} \right) \right|.
\end{align}
The orthonormality condition in the left wedge reads
\begin{align}\label{many_particle_Rindler_basis_product_L}
& \left\langle  \mathcal{O} \left( \bigcup_{i=1}^n \{ (\text{L}, \theta_i) \} \right) \left|  \mathcal{O} \left( \bigcup_{i=1}^m \{ (\text{L}, \theta'_i) \} \right) \right. \right\rangle \nonumber \\
 = & \delta_{nm}  \sum_{\tau \in S_n} \prod_{i = 1}^n \Delta \left( \left( \text{L}, \theta_{\tau (i)} \right), \left(\text{L}, \theta'_i \right) \right).
\end{align}
By plugging Eq.~(\ref{many_particle_Rindler_basis_product_L}) in Eq.~(\ref{tr_L_exp_O}) and by computing the sum $\sum_{m=1}^\infty$ and the generalized sums $\sum_{\theta'_1} \dots \sum_{\theta'_m}$, one obtains
\begin{align}\label{tr_L_exp_O_2}
& \text{Tr}_\text{L} \left[ \exp(\hat{O}) | 0_\text{L},0_\text{R} \rangle \langle 0_\text{L},0_\text{R} | \exp(\hat{O})^\dagger \right]  = |0_\text{R} \rangle \langle 0_\text{R} | \nonumber \\
 & +  \sum_{n=1}^\infty \frac{1}{(n!)^2}  \sum_{\tau \in S_n} \sum_{\theta_1} \dots \sum_{\theta_n} \prod_{i=1}^n f(\theta_i)  \prod_{i=1}^n f^*(\theta_{\tau (i)})   \nonumber \\
& \times \left| \mathcal{O} \left( \bigcup_{i=1}^n \{ (\text{R}, \theta_i) \} \right) \right\rangle \left\langle \mathcal{O} \left( \bigcup_{i=1}^n \{ (\text{R}, \theta_{\tau (i)}) \} \right) \right|.
\end{align}
By using Eq.~(\ref{O_ordering_permutation}) and the fact that the cardinality of $S_n$ is $ n!$, Eq.~(\ref{tr_L_exp_O_2}) reads
\begin{align}\label{tr_L_exp_O_3}
& \text{Tr}_\text{L} \left[ \exp(\hat{O}) | 0_\text{L},0_\text{R} \rangle \langle 0_\text{L},0_\text{R} | \exp(\hat{O})^\dagger \right]  \nonumber \\
=  &  |0_\text{R} \rangle \langle 0_\text{R} | +  \sum_{n=1}^\infty \frac{1}{n!}  \sum_{\theta_1} \dots \sum_{\theta_n} \prod_{i=1}^n |f(\theta_i)|^2 \nonumber \\
& \times \left| \mathcal{O} \left( \bigcup_{i=1}^n \{ (\text{R}, \theta_i) \} \right) \right\rangle \left\langle \mathcal{O} \left( \bigcup_{i=1}^n \{ (\text{R}, \theta_i) \} \right) \right|.
\end{align}

The right side of Eq.~(\ref{tr_L_exp_O_3}) is proportional to the thermal state in the right wedge. This can be seen by considering the following eigenstate decomposition of the Hamiltonian operator
\begin{align}\label{H_R}
& \hat{H}_\text{R} = \sum_{n=1}^\infty \frac{1}{n!}  \sum_{\theta_1} \dots \sum_{\theta_n} \left[ \sum_{i=1}^n h_\text{R}(\theta_i) \right] \nonumber \\
& \times \left| \mathcal{O} \left( \bigcup_{i=1}^n \{ (\text{R}, \theta_i) \} \right) \right\rangle \left\langle \mathcal{O} \left( \bigcup_{i=1}^n \{ (\text{R}, \theta_i) \} \right) \right|,
\end{align}
where
\begin{equation}\label{h_R}
h_\text{R}(s, \Omega,\vec{K}_\perp) = \hbar |\Omega|.
\end{equation}
The $1/n!$ factor comes from the repetition of any independent $n$ particles state due to the permutation symmetry (\ref{O_ordering_permutation}). Notice that Eqs.~(\ref{tr_L_exp_O_3}) and (\ref{H_R}) have the same eigenstate decomposition but with different eigenvalues. By comparing Eq.~(\ref{f}) with Eq.~(\ref{h_R}), one can derive the following identity relating the eigenvalues of Eqs.~(\ref{tr_L_exp_O_3}) and (\ref{H_R})
\begin{equation}
 \prod_{i=1}^n |f(\theta_i)|^2 = \exp \left( - \frac{\beta}{\hbar} \sum_{i=1}^n h_\text{R}(\theta_i) \right),
\end{equation}
which means that
\begin{equation}\label{tr_L_exp_O_4}
\text{Tr}_\text{L} \left[ \exp(\hat{O}) | 0_\text{L},0_\text{R} \rangle \langle 0_\text{L},0_\text{R} | \exp(\hat{O})^\dagger \right] =  \exp \left( - \frac{\beta}{\hbar}  \hat{H}_\text{R} \right).
\end{equation}

By using Eqs.~(\ref{0_M_0_LR}) and (\ref{tr_L_exp_O_4}) we prove that 
\begin{equation}\label{tr_L_0_M}
\text{Tr}_\text{L} | 0_\text{M} \rangle \langle 0_\text{M} | \propto  \exp \left( - \frac{\beta}{\hbar} \hat{H}_\text{R} \right),
\end{equation}
which is the fermionic thermal state with temperature $\hbar /(k_\text{B} \beta)$, where $k_\text{B}$ is the Boltzmann constant. Equation (\ref{tr_L_0_M}) represents the Minkowski vacuum seen by the accelerated observer with acceleration $c^2 a$.

\section{Spin basis choice}\label{Basis_choice}

The result obtained in Sec.~\ref{Minkowski_vacuum_in_Rindler_spacetime} depends of the basis $\tilde{\mathfrak{W}}_{\nu s}(\Omega, \vec{K}_\perp)$. Indeed, the matrix $ M_{\nu ss'} (\Omega, \vec{K}_\perp)$ appears in the Rindler-Fock representation of the Minkowski vacuum [Eqs.~(\ref{0_M_0_LR}) and (\ref{O})]. From Eq.~(\ref{M}), one can see the relation between $  M_{\nu ss'} (\Omega, \vec{K}_\perp)$ and $\tilde{\mathfrak{W}}_{\nu s}(\Omega, \vec{K}_\perp)$. We find out that different choices for the basis $\tilde{\mathfrak{W}}_{\nu s}(\Omega, \vec{K}_\perp)$ lead to different representations of the Minkowski vacuum in the Rindler spacetime.

In Eq.~(\ref{O}), the matrix $  M_{\nu ss'} (\Omega, \vec{K}_\perp)$ couples modes of one wedge with modes of the opposite wedge. Hence, in the Minkowski vacuum, any solution $\tilde{\mathfrak{W}}_{\text{R} s}(\Omega, \vec{K}_\perp)$ of Eq.~(\ref{W_tilde_constraints}) in the right wedge is coupled with a solution of Eq.~(\ref{W_tilde_constraints}) in the left wedge that is proportional to $\tilde{\mathfrak{W}}_{\text{R} s}(-\Omega, -\vec{K}_\perp)$.

The spin coupling of $| 0_\text{M} \rangle$ is then averaged away by the partial trace over the left wedge in Sec.~\ref{Unruh_effect_for_Dirac_fields}. Indeed, the trace is computed by considering a basis for the left wedge [Eqs.~(\ref{E_CD_a}), (\ref{EE}) and (\ref{many_particle_Rindler_basis})] that absorbs the matrix $  M_{\nu ss'} (\Omega, \vec{K}_\perp)$ in Eq.~(\ref{O_3}) and gives an expression for $| 0_\text{M} \rangle$ without $  M_{\nu ss'} (\Omega, \vec{K}_\perp)$ [Eq.~\ref{O_4}].

Consequently, the result obtained in Sec.~\ref{Unruh_effect_for_Dirac_fields} is independent of the choice for the solutions of Eq.~(\ref{W_tilde_constraints}). Indeed, the thermal state describing the Minkowski vacuum in the right wedge [Eq.~(\ref{tr_L_0_M})] is independent of $\tilde{\mathfrak{W}}_{\nu s}(\Omega, \vec{K}_\perp)$. One can see this by plugging Eqs.~(\ref{E_CD_b}), (\ref{EE}), (\ref{many_particle_Rindler_basis}) and (\ref{H_R}) in Eq.~(\ref{tr_L_0_M}) and noticing that $\tilde{\mathfrak{W}}_{\nu s}(\Omega, \vec{K}_\perp)$ never appears in the explicit form of $\text{Tr}_\text{L} | 0_\text{M} \rangle \langle 0_\text{M} |$.

In this section, we go back to the representation of the Minkowski vacuum in both wedges [Eqs.~(\ref{0_M_0_LR}) and (\ref{O})] and we discuss different choices for the basis $\tilde{\mathfrak{W}}_{\nu s}(\Omega, \vec{K}_\perp)$ that lead to different representations of $| 0_\text{M} \rangle$. We study the operator $\hat{O}$ for different choices of $\tilde{\mathfrak{W}}_{\nu s}(\Omega, \vec{K}_\perp)$ and, hence, for different matrices $M_{\nu ss'} (\Omega, \vec{K}_\perp)$. In other words, we consider different outputs of the function $\hat{O} [ M_{\nu ss'} (\Omega, \vec{K}_\perp)]$.

By looking at Eq.~(\ref{O}), one may conclude that the most natural choice for $\tilde{\mathfrak{W}}_{\nu s}(\Omega, \vec{K}_\perp)$ is such that $M_{\nu ss'} (\Omega, \vec{K}_\perp)$ is proportional to the identity. This choice can be made by adopting any spin basis for the $\nu$ wedge and choosing the spin basis in the other wedge $\bar{\nu}$ such that
\begin{equation}
\tilde{\mathfrak{W}}_{\bar{\nu} s}(\Omega, \vec{K}_\perp) \propto \tilde{\mathfrak{W}}_{\nu s}(-\Omega, \vec{K}_\perp).
\end{equation}
In this way, Eq.~(\ref{M}) reads
\begin{equation}\label{M_propto_identity}
M_{\nu ss'} (\Omega, \vec{K}_\perp) \propto \delta_{ss'}
\end{equation}
and the Minkowski vacuum couples each particle mode of one wedge with the antiparticle mode of same spin number $s$ of the other wedge [Eq.~(\ref{O})].

Possible choices for the unitary matrix $   M_{\nu ss'} (\Omega, \vec{K}_\perp)$ that satisfy Eqs.~(\ref{M_conjugate}) and (\ref{M_propto_identity}) are $\mp \text{sign}(\Omega) i \delta_{ss'}$, and $  \mp s_\nu i \delta_{ss'}$, which, respectively, lead to
\begin{subequations}\label{O_pm_pi2}
\begin{align}
\hat{O} [- \text{sign}(\Omega) i \delta_{ss'}] = &  \sum_{s=1}^2 \int_0^{+\infty} d\Omega \int_{\mathbb{R}^2} d^2 K_\perp e^{-\beta \Omega / 2}  \nonumber \\ 
& \times \left[ \hat{C}^\dagger_{s \text{L}}(\Omega,\vec{K}_\perp)\hat{D}^\dagger_{s \text{R}}(\Omega,-\vec{K}_\perp) \right.  \nonumber \\ 
& \left. - \hat{C}^\dagger_{s \text{R}}(\Omega,\vec{K}_\perp) \hat{D}^\dagger_{s \text{L}}(\Omega,-\vec{K}_\perp)  \right] , \\
\hat{O} [ \text{sign}(\Omega) i \delta_{ss'}] = &  \sum_{s=1}^2 \int_0^{+\infty} d\Omega \int_{\mathbb{R}^2} d^2 K_\perp e^{-\beta \Omega / 2}  \nonumber \\ 
& \times \left[  - \hat{C}^\dagger_{s \text{L}}(\Omega,\vec{K}_\perp) \hat{D}^\dagger_{s \text{R}}(\Omega,-\vec{K}_\perp)  \right.  \nonumber \\ 
& \left. + \hat{C}^\dagger_{s \text{R}}(\Omega,\vec{K}_\perp)\hat{D}^\dagger_{s \text{L}}(\Omega,-\vec{K}_\perp)\right], \\
\hat{O} [- s_\nu i  \delta_{ss'}] = &  \sum_{s=1}^2 \int_0^{+\infty} d\Omega \int_{\mathbb{R}^2} d^2 K_\perp e^{-\beta \Omega / 2}  \nonumber \\ 
& \times \left[ -\hat{C}^\dagger_{s \text{L}}(\Omega,\vec{K}_\perp)\hat{D}^\dagger_{s \text{R}}(\Omega,-\vec{K}_\perp) \right.  \nonumber \\ 
& \left. - \hat{C}^\dagger_{s \text{R}}(\Omega,\vec{K}_\perp) \hat{D}^\dagger_{s \text{L}}(\Omega,-\vec{K}_\perp)  \right] , \\
\hat{O} [ s_\nu i  \delta_{ss'}] = &  \sum_{s=1}^2 \int_0^{+\infty} d\Omega \int_{\mathbb{R}^2} d^2 K_\perp e^{-\beta \Omega / 2}  \nonumber \\ 
& \times \left[  \hat{C}^\dagger_{s \text{L}}(\Omega,\vec{K}_\perp) \hat{D}^\dagger_{s \text{R}}(\Omega,-\vec{K}_\perp)  \right.  \nonumber \\ 
& + \left. \hat{C}^\dagger_{s \text{R}}(\Omega,\vec{K}_\perp)\hat{D}^\dagger_{s \text{L}}(\Omega,-\vec{K}_\perp) \right].
\end{align}
\end{subequations}
By letting the creation operators anticommute [Eq.~(\ref{anticommutating_rules_e})], Eq.~(\ref{O_pm_pi2}) reads
\begin{subequations}\label{O_pm_pi2_2}
\begin{align}
\hat{O}[- \text{sign}(\Omega) i \delta_{ss'}] = &  \sum_{s=1}^2 \int_0^{+\infty} d\Omega \int_{\mathbb{R}^2} d^2 K_\perp e^{-\beta \Omega / 2}  \nonumber \\ 
& \times \left[ \hat{C}^\dagger_{s \text{L}}(\Omega,\vec{K}_\perp)\hat{D}^\dagger_{s \text{R}}(\Omega,-\vec{K}_\perp) \right.  \nonumber \\ 
& \left. + \hat{D}^\dagger_{s \text{L}}(\Omega,-\vec{K}_\perp)\hat{C}^\dagger_{s \text{R}}(\Omega,\vec{K}_\perp)   \right] , \label{O_pm_pi2_a}\\
\hat{O}[\text{sign}(\Omega) i \delta_{ss'}] = &  \sum_{s=1}^2 \int_0^{+\infty} d\Omega \int_{\mathbb{R}^2} d^2 K_\perp e^{-\beta \Omega / 2}  \nonumber \\ 
& \times \left[  \hat{D}^\dagger_{s \text{R}}(\Omega,-\vec{K}_\perp) \hat{C}^\dagger_{s \text{L}}(\Omega,\vec{K}_\perp) \right.  \nonumber \\ 
&  \left.+\hat{C}^\dagger_{s \text{R}}(\Omega,\vec{K}_\perp)\hat{D}^\dagger_{s \text{L}}(\Omega,-\vec{K}_\perp) \right], \label{O_pm_pi2_b} \\
\hat{O} [- s_\nu  i \delta_{ss'}] = &  \sum_{s=1}^2 \int_0^{+\infty} d\Omega \int_{\mathbb{R}^2} d^2 K_\perp e^{-\beta \Omega / 2}  \nonumber \\ 
& \times \left[\hat{D}^\dagger_{s \text{R}}(\Omega,-\vec{K}_\perp) \hat{C}^\dagger_{s \text{L}}(\Omega,\vec{K}_\perp)\right.  \nonumber \\ 
& \left. +  \hat{D}^\dagger_{s \text{L}}(\Omega,-\vec{K}_\perp) \hat{C}^\dagger_{s \text{R}}(\Omega,\vec{K}_\perp) \right] ,\label{O_pm_pi2_c} \\
\hat{O} [ s_\nu i \delta_{ss'}] = &  \sum_{s=1}^2 \int_0^{+\infty} d\Omega \int_{\mathbb{R}^2} d^2 K_\perp e^{-\beta \Omega / 2}  \nonumber \\ 
& \times \left[ \hat{C}^\dagger_{s \text{L}}(\Omega,\vec{K}_\perp) \hat{D}^\dagger_{s \text{R}}(\Omega,-\vec{K}_\perp)  \right.  \nonumber \\ 
& + \left. \hat{C}^\dagger_{s \text{R}}(\Omega,\vec{K}_\perp)\hat{D}^\dagger_{s \text{L}}(\Omega,-\vec{K}_\perp)\right]. \label{O_pm_pi2_d}
\end{align}
\end{subequations}

Notice that result that we obtained for fermionic fields is very similar to the bosonic case. Indeed, the Minkowski vacuum of scalars in Rindler spacetimes is equal to Eq.~(\ref{0_M_0_LR}), but with $\hat{O}$ replaced by the following operator \cite{RevModPhys.80.787, falcone2022non}
\begin{align}\label{Theta_B}
 & \hat{O}_\text{B} =  \int_0^{+\infty} d\Omega \int_{\mathbb{R}^2} d^2 K_\perp e^{-\beta \Omega / 2}  \nonumber \\ 
& \times \left[ \hat{A}^\dagger_\text{L}(\Omega,\vec{K}_\perp)\hat{B}^\dagger_\text{R}(\Omega,-\vec{K}_\perp) + \hat{B}^\dagger_\text{L}(\Omega,-\vec{K}_\perp) \hat{A}^\dagger_\text{R}(\Omega,\vec{K}_\perp ) \right] ,
\end{align}
where $\hat{A}_\nu(\Omega,\vec{K}_\perp)$ and $\hat{B}_\nu(\Omega,\vec{K}_\perp)$ are annihilators of scalar particles and antiparticles. Such operators commute. This means that the order between $\hat{A}_\nu(\Omega,\vec{K}_\perp)$ and $\hat{B}_{\bar{\nu}}(\Omega,-\vec{K}_\perp)$ can be switched to give the following equivalent equations
\begin{subequations}\label{Theta_B_2}
\begin{align}
& \hat{O}_\text{B} =  \int_0^{+\infty} d\Omega \int_{\mathbb{R}^2} d^2 K_\perp e^{-\beta \Omega / 2} \nonumber \\ 
& \times \left[\hat{B}^\dagger_\text{R}(\Omega,-\vec{K}_\perp) \hat{A}^\dagger_\text{L}(\Omega,\vec{K}_\perp) + \hat{A}^\dagger_\text{R}(\Omega,\vec{K}_\perp) \hat{B}^\dagger_\text{L}(\Omega,-\vec{K}_\perp) \right], \label{Theta_B_2_a}\\
 & \hat{O}_\text{B} =  \int_0^{+\infty} d\Omega \int_{\mathbb{R}^2} d^2 K_\perp e^{-\beta \Omega / 2}  \nonumber \\ 
& \times \left[ \hat{B}^\dagger_\text{R}(\Omega,-\vec{K}_\perp) \hat{A}^\dagger_\text{L}(\Omega,\vec{K}_\perp)+ \hat{B}^\dagger_\text{L}(\Omega,-\vec{K}_\perp) \hat{A}^\dagger_\text{R}(\Omega,\vec{K}_\perp ) \right] , \label{Theta_B_2_b} \\
 & \hat{O}_\text{B} =  \int_0^{+\infty} d\Omega \int_{\mathbb{R}^2} d^2 K_\perp e^{-\beta \Omega / 2}  \nonumber \\ 
& \times \left[ \hat{A}^\dagger_\text{L}(\Omega,\vec{K}_\perp)\hat{B}^\dagger_\text{R}(\Omega,-\vec{K}_\perp) + \hat{A}^\dagger_\text{R}(\Omega,\vec{K}_\perp ) \hat{B}^\dagger_\text{L}(\Omega,-\vec{K}_\perp) \right] . \label{Theta_B_2_c}
\end{align}
\end{subequations}

For Dirac fields, such an equivalence does not occur because of the anticommuting property of the creation operators [Eq.~(\ref{anticommutating_rules_e})]. Indeed, any swap between creation operators generates a minus sign. However, any of these minus signs can be canceled out by a change of spin basis. One can see this in Eqs.~(\ref{O_pm_pi2_a}), (\ref{O_pm_pi2_b}), (\ref{O_pm_pi2_c}) and (\ref{O_pm_pi2_d}), which are different representations of $| 0_\text{M} \rangle$ that are equivalent up to a change of spin basis. By comparing Eqs.~(\ref{O_pm_pi2_a}), (\ref{O_pm_pi2_b}), (\ref{O_pm_pi2_c}) and (\ref{O_pm_pi2_d}) with Eqs.~(\ref{Theta_B}), (\ref{Theta_B_2_a}), (\ref{Theta_B_2_b}) and (\ref{Theta_B_2_c}), respectively, one can see a complete analogy between scalar and Dirac fields. 

\section{Conclusions} \label{Conclusions}

We derived the representation of the Minkowski vacuum $| 0_\text{M} \rangle$ in the Rindler spacetime for Dirac fields [Eqs.~(\ref{0_M_0_LR}) and (\ref{O})]. The result is a two modes squeezed state that pairs particle modes of one wedge with antiparticle modes of the other wedge. At variance with the scalar case, the coupling also occurs with respect to the spin number $s$. The coupling matrix $   M_{\nu ss'} (\Omega, \vec{K}_\perp)$ can be diagonalized by suitable choices for the spin basis of the Rindler-Dirac modes [Eq.~(\ref{O_pm_pi2_2})].

By computing the partial trace of $| 0_\text{M} \rangle \langle 0_\text{M} |$ with respect to the left wedge, we derived the statistical operator representing the Minkowski vacuum in the right wedge. This gives a complete description of the state seen by the accelerated observer with acceleration $c^2 a$. The result is a fermionic thermal state $\exp( - \beta \hat{H}_\text{R}/\hbar)$, with $\beta = 2 \pi/(ca)$ and $ \hat{H}_\text{R}$ as the Hamiltonian in the right wedge. The consequent thermal distribution of fermionic particles is at the origin of the Unruh effect for Dirac fields.

\appendix

\section{Bessel functions}\label{Bessel_functions}

This section is dedicated to the modified Bessel function of the second kind $K_\zeta (\xi)$ considered throughout the paper. Here, we use the following integral representation for $K_\zeta (\xi)$ with positive argument \cite{bateman_erdelyi_1955}
\begin{align}\label{Bessel_integral_representation}
& K_\zeta (\xi) = \int_0^\infty d\tau e^{- \xi \cosh(\tau) } \cosh(\zeta \tau), & \xi > 0.
\end{align}
From Eq.~(\ref{Bessel_integral_representation}) it is straightforward to prove Eq.~(\ref{Bessel_conjugate}). At the end of this section, we will also prove Eqs.~(\ref{Bessel_orthonormal}) and (\ref{Bessel_integral_representation_final}).

Alternately to Eq.~(\ref{Bessel_integral_representation}) one may use the following integrals \cite{Oriti}
\begin{equation}
f_\zeta^\pm(\xi) = \frac{1}{2} \int_{\mathbb{R}} d\tau \exp\left( i \xi \sinh(\tau) \pm \zeta \left( - i \frac{\pi}{2} + \tau\right) \right).
\end{equation}
When $\xi > 0$, both functions $f_\zeta^\pm(\xi)$ are a representation of $K_\zeta (\xi)$,
\begin{align}\label{Bessel_integral_representation_2}
& K_\zeta (\xi) = f_\zeta^+(\xi) = f_\zeta^-(\xi), & \xi > 0.
\end{align}
Equation (\ref{Bessel_integral_representation_2}) can be proven by using Eq.~(\ref{Bessel_integral_representation}) and by performing a contour integral of $\exp(- \xi \cosh(\tau) + \zeta \tau) $ with respect to $\tau$ along the rectangle with vertexes $-\infty $, $+\infty$, $+\infty \mp i \pi/2$ and $-\infty \mp i \pi/2$, respectively for $f_\zeta^\pm(\xi)$.

Notice that, for any $\xi > 0$,
\begin{align}\label{Bessel_integral_representation_3_a}
K_\zeta (\xi)  = & \frac{ e^{i \pi \zeta} f_\zeta^+(\xi) - e^{-i \pi \zeta} f_\zeta^-(\xi)}{2 i \sin(\pi \zeta)} \nonumber \\
 = & \frac{1}{2 \sin(\pi \zeta)} \int_{\mathbb{R}} d\tau e^{i \xi \sinh(\tau) } \sin \left( \zeta \left( \frac{\pi}{2} - i \tau  \right) \right)
\end{align}
and that, for any $\xi > 0$,
\begin{align}\label{Bessel_integral_representation_3_b}
0 = &  \frac{  f_\zeta^-(\xi) -  f_\zeta^+(\xi)}{2 i \sin(\pi \zeta)} \nonumber \\
 = & \frac{1}{2 \sin(\pi \zeta)} \int_{\mathbb{R}} d\tau e^{i \xi \sinh(\tau) } \sin \left( \zeta \left( \frac{\pi}{2} + i \tau  \right) \right) \nonumber \\
 = & \frac{1}{2 \sin(\pi \zeta)} \int_{\mathbb{R}} d\tau e^{-i \xi \sinh(\tau) } \sin \left( \zeta \left( \frac{\pi}{2} - i \tau \right) \right).
\end{align}
By considering both Eqs.~(\ref{Bessel_integral_representation_3_a}) and (\ref{Bessel_integral_representation_3_b}) one obtains the following identity that holds for any $\xi \in \mathbb{R}$
\begin{equation}\label{Bessel_integral_representation_4}
\theta(\xi) K_\zeta (\xi)  = \frac{1}{2 \sin(\pi \zeta)} \int_{\mathbb{R}} d\tau e^{i \xi \sinh(\tau) } \sin \left( \zeta \left( \frac{\pi}{2} - i \tau \right) \right).
\end{equation}

Equation (\ref{Bessel_integral_representation_4}) can be used to prove both Eq.~(\ref{Bessel_orthonormal}) and Eq.~(\ref{Bessel_integral_representation_final}). Regarding Eq.~(\ref{Bessel_orthonormal}), the proof reads
\begin{align}
& \int_0^\infty d\xi \left[  K_{-i \zeta-1/2}(\xi) K_{i \zeta'-1/2}(\xi) \right. \nonumber \\
& \left. + K_{i \zeta-1/2}(\xi) K_{-i \zeta'-1/2}(\xi) \right] \nonumber \\
= & \frac{1}{4} \int_{\mathbb{R}} d\tau  \int_{\mathbb{R}} d\tau' \left\lbrace \left[ \sin\left(  -i \pi \zeta - \frac{\pi}{2} \right)   \sin\left(  i \pi \zeta' - \frac{\pi}{2} \right) \right]^{-1}  \right. \nonumber \\
& \times  \sin \left(   -i \frac{\pi \zeta}{2}  - \zeta \tau  -  \frac{\pi}{4} + i \frac{\tau}{2}  \right) \sin \left(  i \frac{\pi \zeta'}{2} + \zeta' \tau' \right. \nonumber \\
& \left. - \frac{\pi}{4} + i \frac{\tau'}{2}  \right) + \left[ \sin\left(  i \pi \zeta - \frac{\pi}{2} \right)   \sin\left( -i \pi \zeta' - \frac{\pi}{2} \right) \right]^{-1} \nonumber \\
& \times  \sin \left(   i \frac{\pi \zeta}{2}  + \zeta \tau  -  \frac{\pi}{4} + i \frac{\tau}{2}  \right) \sin \left( - i \frac{\pi \zeta'}{2} - \zeta' \tau' \right. \nonumber \\
& \left. \left. - \frac{\pi}{4} + i \frac{\tau'}{2}  \right) \right\rbrace \int_{\mathbb{R}} d\xi e^{i \xi [\sinh(\tau) + \sinh(\tau')] } \nonumber \\
= &-\frac{\pi}{8 \cosh ( \pi \zeta )   \cosh (  \pi \zeta' )} \int_{\mathbb{R}} d\tau  \int_{\mathbb{R}} d\tau'\nonumber \\
& \times \left[ \exp \left( \pi \frac{\zeta - \zeta'}{2}   - i (\zeta \tau - \zeta' \tau')  - i \frac{\pi}{2} - \frac{\tau+\tau'}{2}  \right)   \right. \nonumber \\
& + \exp \left(- \pi \frac{\zeta - \zeta'}{2}  + i (\zeta \tau - \zeta' \tau')  + i \frac{\pi}{2} + \frac{\tau+\tau'}{2}  \right)\nonumber \\
&  - \exp \left( \pi \frac{\zeta + \zeta'}{2}   - i (\zeta \tau + \zeta' \tau')  - \frac{\tau-\tau'}{2}  \right)\nonumber \\
&  - \exp \left(-\pi \frac{\zeta + \zeta'}{2}   + i (\zeta \tau + \zeta' \tau')    + \frac{\tau-\tau'}{2}  \right) \nonumber \\
& +  \exp \left( - \pi \frac{\zeta - \zeta'}{2}  + i (\zeta \tau - \zeta' \tau')  - i \frac{\pi}{2}   - \frac{\tau+\tau'}{2}  \right) \nonumber \\
& + \exp \left(\pi \frac{\zeta - \zeta'}{2}  - i (\zeta \tau - \zeta' \tau')  + i \frac{\pi}{2}   + \frac{\tau+\tau'}{2}  \right)\nonumber \\
&  - \exp \left( - \pi \frac{\zeta + \zeta'}{2}    + i (\zeta \tau + \zeta' \tau')  - \frac{\tau-\tau'}{2}  \right)   \nonumber \\
& \left.  - \exp \left(\pi \frac{\zeta + \zeta'}{2}  - i (\zeta \tau + \zeta' \tau')    + \frac{\tau-\tau'}{2}  \right) \right] \nonumber \\
& \times   \delta (\sinh(\tau) + \sinh(\tau')) \nonumber \\
= & \frac{\pi}{8 \cosh ( \pi \zeta )   \cosh (  \pi \zeta' )} \int_{\mathbb{R}} d\tau \frac{1}{\cosh (\tau)} \nonumber \\
& \times \left[  \exp \left( \pi \frac{\zeta + \zeta'}{2}   - i (\zeta - \zeta')\tau   - \tau \right)\right. \nonumber \\
&  + \exp \left(-\pi \frac{\zeta + \zeta'}{2}   + i (\zeta - \zeta' ) \tau   + \tau  \right) \nonumber \\
&  + \exp \left( - \pi \frac{\zeta + \zeta'}{2}    + i (\zeta - \zeta') \tau  - \tau  \right)   \nonumber \\
& \left.  + \exp \left(\pi \frac{\zeta + \zeta'}{2}  - i (\zeta - \zeta' ) \tau    + \tau  \right) \right]\nonumber \\
= & \frac{\pi}{4 \cosh ( \pi \zeta )   \cosh (  \pi \zeta' )}\nonumber \\
& \times \int_{\mathbb{R}} d\tau  \left[  \exp \left( \pi \frac{\zeta + \zeta'}{2}   - i (\zeta - \zeta')\tau  \right)\right. \nonumber \\
& \left. + \exp \left(-\pi \frac{\zeta + \zeta'}{2}   + i (\zeta - \zeta' ) \tau    \right) \right] \nonumber \\
 = &  \pi^2 \frac{e^{\pi \zeta}+e^{-\pi \zeta}}{ 2 \cosh^2(\pi \zeta)}\delta(\zeta-\zeta') \nonumber \\
 = &  \frac{\pi^2 \delta(\zeta-\zeta')}{\cosh(\pi \zeta)}.
\end{align}

Equation (\ref{Bessel_integral_representation_final}), instead, can be proved by using Eq.~(\ref{Bessel_integral_representation_4}) to compute the following Fourier transform
\begin{align}\label{Bessel_integral_representation_5}
& \int_{\mathbb{R}} d\xi \theta(\xi) e^{- i \xi \sinh(\tau)} K_\zeta (\xi) \nonumber \\
  = &  \int_{\mathbb{R}} d\tau' \int_{\mathbb{R}} d\xi \frac{e^{i \xi [\sinh(\tau') - \sinh(\tau)] }}{2 \sin(\pi \zeta)} \sin \left( \zeta \left( \frac{\pi}{2} - i \tau' \right) \right) \nonumber \\
  = &  \frac{\pi }{ \sin(\pi \zeta)} \int_{\mathbb{R}} d\tau' \delta (\sinh(\tau') - \sinh(\tau)) \sin \left( \zeta \left( \frac{\pi}{2} - i \tau' \right) \right) \nonumber \\
  = &  \frac{\pi}{ \sin(\pi \zeta) \cosh(\tau)} \sin \left( \zeta \left( \frac{\pi}{2} - i \tau \right) \right).
\end{align}

\bibliography{bibliography}

\begin{thebibliography}{11}%
\makeatletter
\providecommand \@ifxundefined [1]{%
 \@ifx{#1\undefined}
}%
\providecommand \@ifnum [1]{%
 \ifnum #1\expandafter \@firstoftwo
 \else \expandafter \@secondoftwo
 \fi
}%
\providecommand \@ifx [1]{%
 \ifx #1\expandafter \@firstoftwo
 \else \expandafter \@secondoftwo
 \fi
}%
\providecommand \natexlab [1]{#1}%
\providecommand \enquote  [1]{``#1''}%
\providecommand \bibnamefont  [1]{#1}%
\providecommand \bibfnamefont [1]{#1}%
\providecommand \citenamefont [1]{#1}%
\providecommand \href@noop [0]{\@secondoftwo}%
\providecommand \href [0]{\begingroup \@sanitize@url \@href}%
\providecommand \@href[1]{\@@startlink{#1}\@@href}%
\providecommand \@@href[1]{\endgroup#1\@@endlink}%
\providecommand \@sanitize@url [0]{\catcode `\\12\catcode `\$12\catcode
  `\&12\catcode `\#12\catcode `\^12\catcode `\_12\catcode `\%12\relax}%
\providecommand \@@startlink[1]{}%
\providecommand \@@endlink[0]{}%
\providecommand \url  [0]{\begingroup\@sanitize@url \@url }%
\providecommand \@url [1]{\endgroup\@href {#1}{\urlprefix }}%
\providecommand \urlprefix  [0]{URL }%
\providecommand \Eprint [0]{\href }%
\providecommand \doibase [0]{https://doi.org/}%
\providecommand \selectlanguage [0]{\@gobble}%
\providecommand \bibinfo  [0]{\@secondoftwo}%
\providecommand \bibfield  [0]{\@secondoftwo}%
\providecommand \translation [1]{[#1]}%
\providecommand \BibitemOpen [0]{}%
\providecommand \bibitemStop [0]{}%
\providecommand \bibitemNoStop [0]{.\EOS\space}%
\providecommand \EOS [0]{\spacefactor3000\relax}%
\providecommand \BibitemShut  [1]{\csname bibitem#1\endcsname}%
\let\auto@bib@innerbib\@empty
\bibitem [{\citenamefont {Fulling}(1973)}]{PhysRevD.7.2850}%
  \BibitemOpen
  \bibfield  {author} {\bibinfo {author} {\bibfnamefont {S.~A.}\ \bibnamefont
  {Fulling}},\ }\bibfield  {title} {\bibinfo {title} {{Nonuniqueness of
  Canonical Field Quantization in Riemannian Space-Time}},\ }\href
  {https://doi.org/10.1103/PhysRevD.7.2850} {\bibfield  {journal} {\bibinfo
  {journal} {Phys. Rev. D}\ }\textbf {\bibinfo {volume} {7}},\ \bibinfo {pages}
  {2850} (\bibinfo {year} {1973})}\BibitemShut {NoStop}%
\bibitem [{\citenamefont {Davies}(1975)}]{Davies:1974th}%
  \BibitemOpen
  \bibfield  {author} {\bibinfo {author} {\bibfnamefont {P.~C.~W.}\
  \bibnamefont {Davies}},\ }\bibfield  {title} {\bibinfo {title} {{Scalar
  particle production in Schwarzschild and Rindler metrics}},\ }\href
  {https://doi.org/10.1088/0305-4470/8/4/022} {\bibfield  {journal} {\bibinfo
  {journal} {J. Phys. A}\ }\textbf {\bibinfo {volume} {8}},\ \bibinfo {pages}
  {609} (\bibinfo {year} {1975})}\BibitemShut {NoStop}%
\bibitem [{\citenamefont {Unruh}(1976)}]{PhysRevD.14.870}%
  \BibitemOpen
  \bibfield  {author} {\bibinfo {author} {\bibfnamefont {W.~G.}\ \bibnamefont
  {Unruh}},\ }\bibfield  {title} {\bibinfo {title} {Notes on black-hole
  evaporation},\ }\href {https://doi.org/10.1103/PhysRevD.14.870} {\bibfield
  {journal} {\bibinfo  {journal} {Phys. Rev. D}\ }\textbf {\bibinfo {volume}
  {14}},\ \bibinfo {pages} {870} (\bibinfo {year} {1976})}\BibitemShut
  {NoStop}%
\bibitem [{\citenamefont {Oriti}(1999)}]{Oriti}%
  \BibitemOpen
  \bibfield  {author} {\bibinfo {author} {\bibfnamefont {D.}~\bibnamefont
  {Oriti}},\ }\bibfield  {title} {\bibinfo {title} {{The spinor field in
  Rindler spacetime: An analysis of the Unruh effect}},\ }\href@noop {}
  {\bibfield  {journal} {\bibinfo  {journal} {Nuovo Cimento della Societa
  Italiana di Fisica B}\ }\textbf {\bibinfo {volume} {115}} (\bibinfo {year}
  {1999})}\BibitemShut {NoStop}%
\bibitem [{\citenamefont {Ueda}\ \emph {et~al.}(2021)\citenamefont {Ueda},
  \citenamefont {Higuchi}, \citenamefont {Yamamoto}, \citenamefont {Rohim},\
  and\ \citenamefont {Nan}}]{PhysRevD.103.125005}%
  \BibitemOpen
  \bibfield  {author} {\bibinfo {author} {\bibfnamefont {K.}~\bibnamefont
  {Ueda}}, \bibinfo {author} {\bibfnamefont {A.}~\bibnamefont {Higuchi}},
  \bibinfo {author} {\bibfnamefont {K.}~\bibnamefont {Yamamoto}}, \bibinfo
  {author} {\bibfnamefont {A.}~\bibnamefont {Rohim}},\ and\ \bibinfo {author}
  {\bibfnamefont {Y.}~\bibnamefont {Nan}},\ }\bibfield  {title} {\bibinfo
  {title} {{Entanglement of the vacuum between left, right, future, and past:
  Dirac spinor in Rindler and Kasner spaces}},\ }\href
  {https://doi.org/10.1103/PhysRevD.103.125005} {\bibfield  {journal} {\bibinfo
   {journal} {Phys. Rev. D}\ }\textbf {\bibinfo {volume} {103}},\ \bibinfo
  {pages} {125005} (\bibinfo {year} {2021})}\BibitemShut {NoStop}%
\bibitem [{\citenamefont {Wald}(1994)}]{wald1994quantum}%
  \BibitemOpen
  \bibfield  {author} {\bibinfo {author} {\bibfnamefont {R.}~\bibnamefont
  {Wald}},\ }\bibinfo {title} {{Quantum Field Theory in Curved Spacetime and
  Black Hole Thermodynamics}}\ (\bibinfo  {publisher} {University of Chicago
  Press},\ \bibinfo {year} {1994})\ Chap.\ \bibinfo {chapter} {4.5}\BibitemShut
  {NoStop}%
\bibitem [{\citenamefont {Falcone}\ and\ \citenamefont
  {Conti}(2023{\natexlab{a}})}]{falcone2022non}%
  \BibitemOpen
  \bibfield  {author} {\bibinfo {author} {\bibfnamefont {R.}~\bibnamefont
  {Falcone}}\ and\ \bibinfo {author} {\bibfnamefont {C.}~\bibnamefont
  {Conti}},\ }\bibfield  {title} {\bibinfo {title} {{Nonrelativistic limit of
  scalar and Dirac fields in curved spacetime}},\ }\href
  {https://doi.org/10.1103/PhysRevD.107.045012} {\bibfield  {journal} {\bibinfo
   {journal} {Phys. Rev. D}\ }\textbf {\bibinfo {volume} {107}},\ \bibinfo
  {pages} {045012} (\bibinfo {year} {2023}{\natexlab{a}})}\BibitemShut
  {NoStop}%
\bibitem [{\citenamefont {Abramowitz}\ and\ \citenamefont
  {Stegun}(1965)}]{abramowitz1965handbook}%
  \BibitemOpen
  \bibfield  {author} {\bibinfo {author} {\bibfnamefont {M.}~\bibnamefont
  {Abramowitz}}\ and\ \bibinfo {author} {\bibfnamefont {I.}~\bibnamefont
  {Stegun}},\ }\href {https://books.google.it/books?id=MtU8uP7XMvoC} {\emph
  {\bibinfo {title} {{Handbook of Mathematical Functions: With Formulas,
  Graphs, and Mathematical Tables}}}},\ Applied mathematics series\ (\bibinfo
  {publisher} {Dover Publications},\ \bibinfo {year} {1965})\BibitemShut
  {NoStop}%
\bibitem [{\citenamefont {Falcone}\ and\ \citenamefont
  {Conti}(2023{\natexlab{b}})}]{falcone2022}%
  \BibitemOpen
  \bibfield  {author} {\bibinfo {author} {\bibfnamefont {R.}~\bibnamefont
  {Falcone}}\ and\ \bibinfo {author} {\bibfnamefont {C.}~\bibnamefont
  {Conti}},\ }\href@noop {} {\bibinfo {title} {Frame-dependence of the
  non-relativistic limit of quantum fields}} (\bibinfo {year}
  {2023}{\natexlab{b}}),\ \Eprint {https://arxiv.org/abs/2301.13011}
  {arXiv:2301.13011 [hep-th]} \BibitemShut {NoStop}%
\bibitem [{\citenamefont {Crispino}\ \emph {et~al.}(2008)\citenamefont
  {Crispino}, \citenamefont {Higuchi},\ and\ \citenamefont
  {Matsas}}]{RevModPhys.80.787}%
  \BibitemOpen
  \bibfield  {author} {\bibinfo {author} {\bibfnamefont {L.~C.~B.}\
  \bibnamefont {Crispino}}, \bibinfo {author} {\bibfnamefont {A.}~\bibnamefont
  {Higuchi}},\ and\ \bibinfo {author} {\bibfnamefont {G.~E.~A.}\ \bibnamefont
  {Matsas}},\ }\bibfield  {title} {\bibinfo {title} {{The Unruh effect and its
  applications}},\ }\href {https://doi.org/10.1103/RevModPhys.80.787}
  {\bibfield  {journal} {\bibinfo  {journal} {Rev. Mod. Phys.}\ }\textbf
  {\bibinfo {volume} {80}},\ \bibinfo {pages} {787} (\bibinfo {year}
  {2008})}\BibitemShut {NoStop}%
\bibitem [{\citenamefont {Erdélyi}\ \emph {et~al.}(1955)\citenamefont
  {Erdélyi}, \citenamefont {Magnus}, \citenamefont {Oberhettinger},\ and\
  \citenamefont {Tricomi}}]{bateman_erdelyi_1955}%
  \BibitemOpen
  \bibfield  {author} {\bibinfo {author} {\bibfnamefont {A.}~\bibnamefont
  {Erdélyi}}, \bibinfo {author} {\bibfnamefont {W.}~\bibnamefont {Magnus}},
  \bibinfo {author} {\bibfnamefont {F.}~\bibnamefont {Oberhettinger}},\ and\
  \bibinfo {author} {\bibfnamefont {F.~G.}\ \bibnamefont {Tricomi}},\
  }\href@noop {} {\emph {\bibinfo {title} {{Higher Transcendental Functions.
  Vol. II}}}}\ (\bibinfo  {publisher} {McGraw-Hill Book Company, Inc.},\
  \bibinfo {year} {1955})\BibitemShut {NoStop}%
\end{thebibliography}%

\end{document}